\documentclass[fleqn,usenatbib]{mnras}

\usepackage{newtxtext,newtxmath}

\usepackage[T1]{fontenc}
\usepackage{ae,aecompl}
\usepackage{siunitx}
\usepackage[para]{threeparttable}
\usepackage{longtable}
\usepackage{tabularray}
\usepackage{subfigure}

\usepackage{graphicx}	
\usepackage{amsmath}	
\usepackage{wrapfig,lipsum}
\usepackage{hyperref}
\usepackage{upgreek}

\newcommand{\FRB}{FRB\,20240114A}

\title[FRB search]{Detection and localisation of the highly active \FRB\ with MeerKAT}

\author[J. Tian et al.]{J. Tian$^{1}$\thanks{E-mail: jun.tian@manchester.ac.uk}, K. M. Rajwade$^2$, I. Pastor-Marazuela$^1$, B. W. Stappers$^1$, M. C. Bezuidenhout$^{3,4}$,
\newauthor M. Caleb$^{5,6}$, F. Jankowski$^7$, E. D. Barr$^{8}$, M. Kramer$^8$
\\
$^1$Jodrell Bank Centre for Astrophysics, Department of Physics and Astronomy, The University of Manchester, Manchester M13 9PL, UK\\
$^2$Astrophysics, The University of Oxford, Denys Wilkinson Building, Keble Road,
Oxford OX1 3RH, UK\\
$^3$Centre for Space Research, North-West University, Potchefstroom 2531, South Africa\\
$^{4}$Department of Mathematical Sciences, University of South Africa, Cnr Christiaan de Wet Rd and Pioneer Avenue, Florida Park, 1709, Roodepoort, South Africa\\
$^5$Sydney Institute for Astronomy, School of Physics, The University of Sydney, NSW 2006, Australia\\
$^6$ARC Centre of Excellence for Gravitational Wave Discovery (OzGrav), Hawthorn, 3122, Victoria, Australia\\
$^7$LPC2E, Universite d’Orleans, CNRS, 3A Avenue de la Recherche Scientifique, F-45071 Orleans, France\\
$^8$Max-Planck-Institut fur Radioastronomie, 53121 Bonn, Germany
}%

\date{Accepted XXX. Received YYY; in original form ZZZ}

\pubyear{2024}

\begin{document}
\label{firstpage}
\pagerange{\pageref{firstpage}--\pageref{lastpage}}
\maketitle

\begin{abstract}
We report observations of the highly active \FRB\ with MeerKAT using the Ultra-High Frequency (UHF; $544\text{--}1088$\,MHz) and L-band ($856\text{--}1712$\,MHz) receivers. A total of 62 bursts were detected in coherent tied-array beams using the MeerTRAP real-time transient detection pipeline. We measure a structure-optimising dispersion measure of $527.65\pm0.01\,\text{pc}\,\text{cm}^{-3}$ using the brightest burst in the sample. 
We find the bursts of \FRB\ are generally detected in part of the broad band of MeerKAT, $\sim40\%$ in the UHF and $\sim30\%$ in the L-band, indicating the band limited nature.
We analyse the fluence distribution of the 44 bursts detected at UHF, constraining the fluence completeness limit to $\sim1$\,Jy\,ms, above which the cumulative burst rate follows a power law $R (>F)\propto (F/1\,\text{Jy}\,\text{ms})^\gamma$ with $\gamma=-1.8\pm0.2$. Using channelised telescope data captured in our transient buffer we localise \FRB\ in the image domain to RA = 21h27m39.86s, Dec = +04d19m45.01s with an uncertainty of 1.4\,arcsec. This localisation allows us to confidently identify the host galaxy of \FRB. Also using the transient buffer data we perform a polarimetric study and demonstrate that most of the bursts have $\sim100\%$ linear polarisation fractions and up to $\sim20\%$ circular polarisation fractions. Finally, we predict the flux density of a potential persistent radio source (PRS) associated with \FRB\ is $\backsimeq[0.6\text{--}60]\,\mu\text{Jy}$ based on the simple relation between the luminosity of the PRS and the rotation measure arising from the FRB local environment. 
\end{abstract}




\section{INTRODUCTION}

Fast radio bursts (FRBs) are bright ($\sim10\,\text{mJy}\text{--}100\,\text{Jy}$; e.g. \citealt{Petroff16, Shannon18}), short-duration ($\mu\text{s}\text{--}\text{ms}$; e.g. \citealt{Cho20, Nimmo21}) bursts of radio emission with extra-galactic origins, mainly hosted by star-forming galaxies \citep{Gordon23}. The last few years have witnessed a rapid expansion of the FRB population, with most discovered by the Canadian Hydrogen Intensity Mapping Experiment (CHIME; \citealt{CHIME22}) Fast Radio Burst project (CHIME/FRB; \citealt{CHIME18}). However, the physical origin of FRBs remains unknown although several lines of evidence favor a neutron star origin (e.g. \citealt{Zhang23}).

While most FRBs appear to be one-off events, a subset ($\sim50$ known) are observed to repeat \citep{Chime23}, allowing detailed studies of these FRB sources with targeted follow-up observations. Collecting a large sample of bursts from a repeating FRB source can reveal important features in the burst properties and progress our understanding of the burst emission mechanism and the influence of propagation effects, such as the downward drifting of subpulses in frequency with time \citep{Hessels19, CHIME19c, Caleb20, Pleunis21b} and the evolution of polarisation position angles with time \citep{Michilli18, Luo20, Nimmo21}. In particular, measuring the polarisation fractions and rotation measures (RMs) of a large sample of bursts could reveal any depolarisation towards lower frequencies and RM scattering, which are intriguing features of repeating FRBs that have started to emerge recently and shed light on the complexity of magnetized environments associated with repeating FRBs \citep{Feng22, Anna-Thomas23}. 

Repeating FRBs also allow for periodicity searches. On long timescales ($\sim$days), there is evidence for periodic activity from FRB 20121102A \citep{Rajwade20c, Cruces21} and FRB 20180916B \citep{CHIME20b}, which might reflect an orbital \citep{Ioka20}, rotational \citep{Beniamini20} or precession \citep{Levin20} period. On short timescales ($\sim1\,\text{ms}\,\text{--}\,1\,\text{s}$), quasi-periodic substructures have been reported for FRB 20200120E \citep{Majid21}, FRB 20191221A \citep{CHIME22b} and FRB 20201020A \citep{Pastor-Marazuela23}. Such substructures resemble the quasi-periodic micropulses seen in some radio pulsars \citep{Mitra15, De16} and magnetars \citep{Kramer24}, favoring a neutron star origin for FRBs. These motivate further searches for periodicities in repeating FRBs, especially in the range of $\sim1\,\text{s}\,\text{--}\,1000\,\text{s}$, which could indicate associations with magnetars.

Apart from studies of the emission properties of repeating FRBs, their repeating nature facilitates accurate localisation of the source. The first subarcsecond localisation of an FRB was made through targeted observations of FRB 20121102A, associating this repeater with a low-metallicity star-forming dwarf galaxy \citep{Chatterjee17, Tendulkar17}. Later, more subarcsecond localisations of repeating FRBs revealed a wide range of host galaxies, including a massive spiral galaxy \citep{Marcote20}, a star-forming, dusty and massive galaxy similar to that of apparent non-repeaters \citep{Ravi22}, a globular cluster \citep{Kirsten22}, and an additional dwarf galaxy \citep{Niu22}. Such a diversity of host galaxies imply that multiple progenitors may be responsible for the FRB phenomenon, motivating the characterisation of a larger sample, which could provide further insight into the origin of FRB repetition \citep{Gordon23}.

\FRB\ is a repeating source recently discovered by CHIME/FRB at a dispersion measure (DM) of $527.7\,\text{pc}\,\text{cm}^{-3}$ and reported to be in an active state from the end of January 2024 \citep{Shin24}. It was shown to be highly energetic with the brightest burst reaching a fluence of $919\pm97$\,Jy\,ms. An initial localisation of the source by CHIME/FRB gave RA = 21:27:39.89, Dec = +04:21:00.36 with a $\sim30$\,arcsec uncertainty. These motivated us to propose for Director's Discretionary Time to follow up with MeerKAT (proposal id: DDT-20240206-JT-01), which was allocated 2\,hr of observation on 2024 February 9. Subsequent detections of many more bursts by Murriyang \citep{Uttarkar24}, the Five-hundred-meter Aperture Spherical radio Telescope (FAST; \citealt{Zhang24}) and the upgraded Giant Metrewave Radio Telescope (uGMRT; \citealt{Kumar24, Panda24}) confirmed the hyperactivity of \FRB. Our observation with MeerKAT improved the source position to $\sim$arcsec precision and made the first identification of the host galaxy \citep{Tian24}, which were then used by uGMRT and FAST in their targeted searches, leading to the detections of hundreds of bursts \citep{Panda24, Zhang24b}. Later, the European Very Long Baseline Interferometry (VLBI) PRECISE team localised \FRB\ to RA = 21:27:39.835, Dec = +04:19:45.634 with an uncertainty of 200\,milliarcseconds \citep{Snelders24}, which is consistent with our localisation.

In this paper we describe the detection of 62 bursts from \FRB\ with MeerKAT and present the burst properties and localisation of the source. In Section 2, we describe the observational configuration of MeerKAT and the transient detection pipeline. Our results are then presented in Section 3. We discuss the host galaxy of \FRB\ and inferred properties in Section 4, followed by conclusions in Section 5.

\section{Observations and data}\label{sec:observation}

The MeerKAT observations of \FRB\ were carried out on 2024 February 09 as part of the DDT proposal DDT-20240206-JT-01 with the Ultra-High Frequency (UHF; 544--1088\,MHz) and L-band (856--1712\,MHz) receivers on 40 of the 64 13.5-m dishes in the inner $\sim$1-km core of the array. The observations lasted 2\,hr with 1\,hr at UHF and 1\,hr at L-band. The primary beam full width at half-maximum (FWHM) at the UHF and L-band are $\sim3.2\,\text{deg}^2$ and $\sim1.3\,\text{deg}^2$, respectively \citep{Mauch20}. At that time, \FRB\ was localised to a $\sim30$\,arcsec region by CHIME/FRB, well within the primary beam of MeerKAT. Therefore, we recorded imaging data for the FRB source localisation as well as beamformed data to search for repeat bursts.

Thanks to the Transient User Supplied Equipment (TUSE), a real-time transient detection backend instrument developed by the Meer(more) TRAnsients and Pulsars (MeerTRAP; \citealt{Sanidas18, Bezuidenhout22, Rajwade22, Caleb23, Jankowski23, Driessen24}) project, we are able to trigger voltage buffer dumps while searching for bursts in real-time. 
Our follow-up of \FRB\ started with the coherent beamforming mode, where voltages from the inner 40 dishes were coherently combined and phased using the Filterbank and Beamforming User Supplied Equipment (FBFUSE), a many-beam beamformer that was designed and developed at the Max Planck Institute for Radio Astronomy in Bonn \citep{Barr18, Chen21}. We formed 768 tied-array coherent beams (CBs) overlapping at 75\% of the beam power that tiled out from the best localisation of \FRB\ reported by CHIME/FRB up to a radius of $\sim3$\,arcmin. Data from these CBs were arranged and ingested by the TUSE real-time single pulse search pipeline, enabling us to instantaneously detect multiple pulses from FRB~20240114A and providing a quicker initial investigation. In the case of a detection, channelised, high time resolution transient buffer data were saved for offline correlation and imaging (for the MeerTRAP voltage buffer dump system see \citealt{Rajwade24}). Note that the triggering system is limited by the processing speed of the real-time pulse search and also bursts within 10\,s of each other are not triggered to limit the load on the system so we did not get voltage capture triggers for all bursts.

Given the above observing strategy, our observation of \FRB\ resulted in two datasets: detected time-frequency data in the filterbank format from all CB detections and voltage data for all triggers. Each filterbank file contains a dispersed pulse and additional padding of 0.5\,s at the start and end of the file. The full bandwidth is split into 1024 channels, corresponding to a frequency and time resolution of 0.53\,MHz/481.88\,$\mu$s and 0.84\,MHz/306.24\,$\mu$s at UHF and L-band, respectively. For the subset of the detections that triggered the transient buffer, we additionally recorded $\sim300$\,ms channelised voltage data, which had been incoherently dedispersed at the detection DM. These voltage data are Nyquist sampled across 4096 channels from $\sim60$ out of 64 dishes available at that time. 

\subsection{Burst detection}

\begin{figure*}
\centering
\includegraphics[width=.85\textwidth]{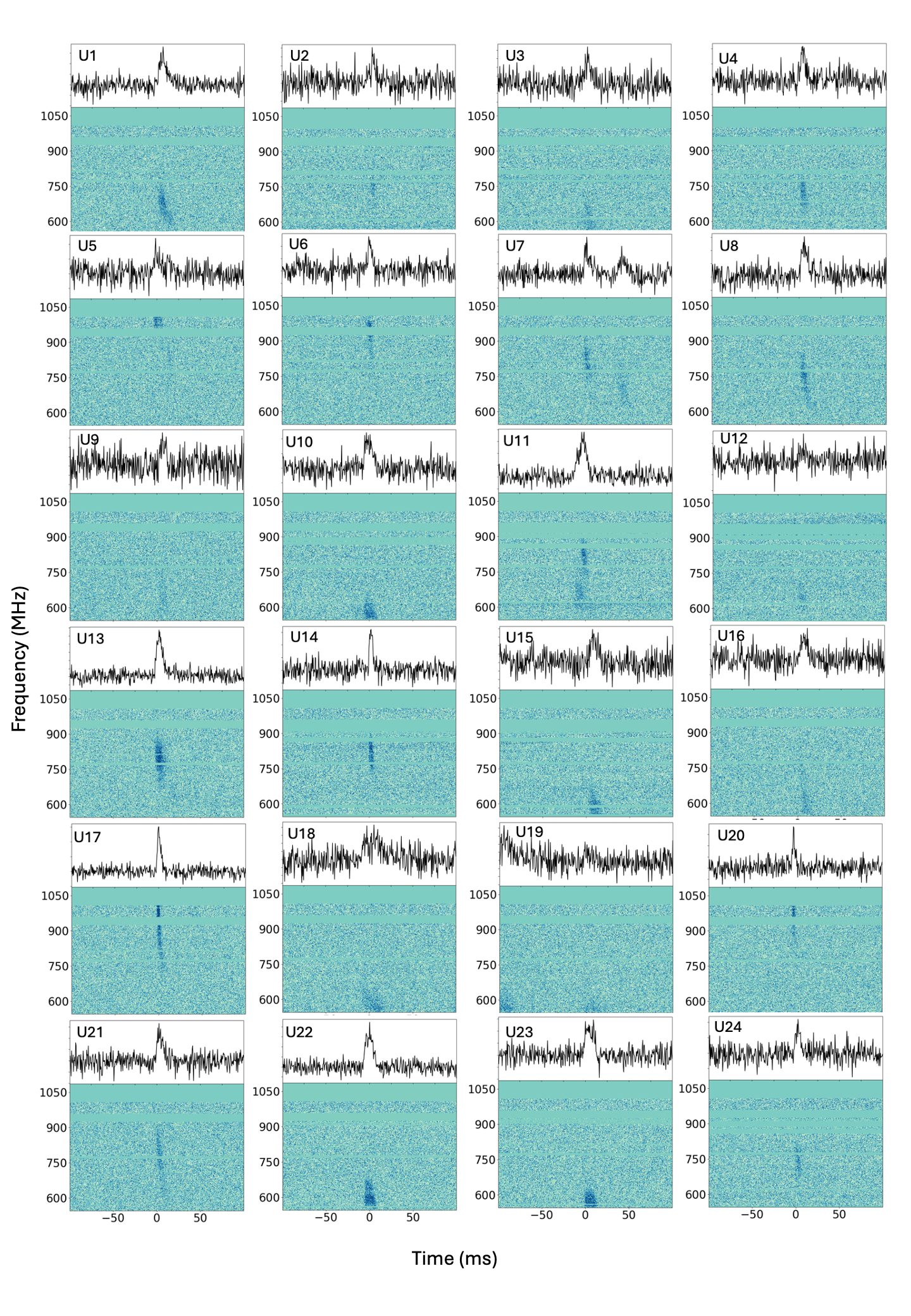}
\caption{Dynamic spectra of the bursts detected from \FRB\ in chronological order. Each panel shows the dynamic spectrum of a burst from Table~\ref{tab:bursts} from the filterbank data dedispersed to $527.65\,\text{pc}\,\text{cm}^{-3}$, the structure optimising DM determined for \FRB\ in Section~\ref{sec:DM}, with the top sub-panel showing the frequency-averaged pulse profile in arbitrary units. The dynamic spectra have been binned $4\times$ in frequency. A label is given to each burst in the top-left corner with "U" and "L" indicating detection in the UHF and L-band, respectively. Burst U19 follows U18 closely, and part of the U18 emission is visible in the panel of U19. Bursts U43 and U44 were detected only $\sim50$\,ms apart and are presented in the same panel. So are bursts L9 and L10. The horizontal lines that show the same colour are either missing channels or flagged due to RFI.}
\label{fig:bursts}
\end{figure*}

We performed a real-time burst search on each CB targeted at \FRB\ using the state-of-the-art GPU-based single pulse search pipeline {\sc astroaccelerate}\footnote{\url{https://github.com/AstroAccelerateOrg/astro-accelerate}} \citep{Armour12, Adamek20}. All candidates with signal-to-noise (S/N) above 8 were saved to disc in the filterbank format. See \citealt{Caleb22}, \citealt{Rajwade22} and \citealt{Jankowski23} for more details on removing radio frequency interference (RFI) and sifting the candidates. After visually inspecting the pulse profiles and dynamic spectra of the candidates and removing any repetitions of the same burst detected in multiple CBs, 
we obtained a list of 62 bursts with detection DMs ranging from $527.84\,\text{pc}\,\text{cm}^{-3}$ to $534.36\,\text{pc}\,\text{cm}^{-3}$, 44 in the UHF and 18 in the L-band over the full exposure time of 2\,hr.
The gallery of the detected bursts dedispersed to $527.65\,\text{pc}\,\text{cm}^{-3}$ (see Section~\ref{sec:DM}) is shown in Figure~\ref{fig:bursts}, with their times of arrival (TOAs, barycentric and referenced to infinite frequency), detected S/Ns, burst widths and fluences being listed in Table~\ref{tab:bursts}. We cover $\pm100$\,ms around the time of the burst.

\begin{table*}
\begin{threeparttable}
\centering
\resizebox{1.5\columnwidth}{!}{\hspace{-0cm}\begin{tabular}
{l c r r c c c c c}
\hline
 Burst\tnote{a} & TOA\tnote{b} & $\text{S/N}_\text{det}$\tnote{c} & Burst width\tnote{d} & Fluence\tnote{e} & Error\tnote{e} & Trigger\tnote{f} & $L/I$\tnote{g} & $|V|/I$\tnote{g} \\
  & (MJD)&  & (ms) & (Jy ms) & (Jy ms) & & & \\
\hline
 U1 & 60349.25540534 & 22.0 & 8.8 & 2.44 & 0.26 & Y & 0.98(13) & 0.16(6) \\
 U2 & 60349.25567773 & 8.8 & 5.6 & 1.87 & 0.20 & N & & \\
 U3 & 60349.25568583 & 11.1 & 8.8 & 3.68 & 0.31 & N & & \\
 U4 & 60349.25576426 & 12.9 & 5.6 & 2.28 & 0.24 & Y & 0.99(15) & 0.16(7) \\
 U5 & 60349.25687076 & 12.8 & 11.0 & 2.90 & 0.12 & Y & 0.94(18) & 0.18(10) \\
 U6 & 60349.25922901 & 16.1 & 4.5 & 1.12 & 0.09 & Y & 0.79(17) & 0.11(8)\\
 U7 & 60349.25963037 & 18.4 & 5.6 & 1.60 & 0.18 & Y & 1.26(50) & 0.20(22) \\
 U8 & 60349.25978003 & 21.0 & 5.6 & 1.20 & 0.09 & Y & 1.08(16) & 0.22(8) \\
 U9 & 60349.25978165 & 8.9 & 7.0 & 0.63 & 0.11 & N & & \\
 U10 &60349.26020843 & 16.4 & 8.8 & 3.64 & 0.30 & Y & 0.88(17) &0.11(6) \\
 U11 &60349.26112171 & 30.2 & 8.8 & 2.61 & 0.26 & Y & 0.80(9) & 0.41(6) \\
 U12$^\ast$ &60349.26118684 & 6.7 & 1.8 & 0.12 & 0.05 & N & & \\
 U13 &60349.26584730 & 37.1 & 8.8 & 2.20 & 0.10 & Y &1.05(10) & 0.17(5) \\
 U14 &60349.26916507 & 20.9 & 3.6 & 2.25 & 0.15 & Y &0.92(14) &0.22(8) \\
 U15 &60349.26918248 & 14.2 & 8.8 & 3.73 & 0.37 & N & & \\
 U16 &60349.27099313 & 12.8 & 7.0 & 1.65 & 0.23 & Y &0.94(27) &0.23(14) \\
 U17 &60349.27108290 & 34.2 & 2.9 & 1.79 & 0.08 & N & & \\
 U18 &60349.27132540 & 13.9 & 21.4 & 2.18 & 0.15 & Y &1.37(47) &0.35(22) \\
 U19$^\ast$ &60349.27132653 & 6.2 & 8.4 & 0.38 & 0.06 & N & & \\
 U20 &60349.27320722 & 14.2 & 2.9 & 3.07 & 0.34 & Y &1.04(17) &0.09(8) \\
 U21 &60349.27355901 & 20.5 & 5.6 & 0.90 & 0.09 & Y &1.10(18) &0.16(8) \\
 U22 &60349.27609791 & 27.5 & 8.8 & 4.01 & 0.22 & Y &0.91(8) & 0.26(4)\\
 U23 &60349.27617901 & 21.3 & 11.0 & 3.34 & 0.20 & N & & \\
 U24 &60349.27657166 & 12.4 & 5.6 & 2.17 & 0.25 & Y &0.99(25) &0.20(11) \\
 U25 &60349.27695634 & 25.4 & 11.0 & 4.88 & 0.34 & Y &0.92(8) &0.13(4) \\
 U26 &60349.27724616 & 17.7 & 7.0 & 1.56 & 0.13 & Y &0.44(8) &0.20(5) \\
 U27 &60349.27784935 & 12.8 & 5.6 & 1.70 & 0.28 & Y &1.06(18) &0.07(8) \\
 U28 &60349.28100674 & 20.2 & 8.8 & 1.15 & 0.10 & Y &0.98(16) &0.24(8) \\
 U29 &60349.28342559 & 177.2 & 3.6 & 7.00 & 0.10 & Y &1.09(10) &0.05(4) \\
 U30 &60349.28552309 & 32.8 & 8.8 & 3.24 & 0.19 & Y &0.92(8) &0.07(4) \\
 U31 &60349.28608044 & 28.1 & 5.6 & 2.04 & 0.13 & Y &0.90(11) &0.30(6) \\
 U32 &60349.28708962 & 9.3 & 5.6 & 0.80 & 0.12 & N & & \\
 U33 &60349.28869337 & 23.8 & 4.5 & 1.36 & 0.08 & Y &0.83(10) &0.14(5) \\
 U34 &60349.28901404 & 21.8 & 5.6 & 1.45 & 0.14 & Y &0.99(16) &0.20(7) \\
 U35 &60349.28921046 & 16.6 & 5.6 & 1.74 & 0.12 & Y &1.06(18) &0.13(9) \\
 U36 &60349.28967727 & 22.6 & 5.6 & 2.10 & 0.14 & Y &0.86(9) &0.06(4) \\
 U37 &60349.29169012 & 87.2 & 7.0 & 7.02 & 0.30 & Y &0.83(5) &0.12(2) \\
 U38 &60349.29212284 & 12.6 & 4.5 & 3.03 & 0.12 & Y &1.33(45) &0.33(22) \\
 U39 &60349.29216244 & 51.8 & 4.5 & 2.11 & 0.11 & N & & \\
 U40 &60349.29217031 & 10.1 & 8.8 & 0.80 & 0.21 & N & & \\
 U41 &60349.29371123 & 16.9 & 5.6 & 1.02 & 0.10 & Y &0.97(12) &0.30(7) \\
 U42 &60349.29569083 & 11.3 & 5.6 & 2.31 & 0.29 & N & & \\
 U43 &60349.29591442 & 15.7 & 8.8 & 1.20 & 0.12 & Y &0.93(9) &0.14(4) \\
 U44$^\ast$ &60349.29591494 & 7.2 & 7.0 & 0.44 & 0.10 & N & & \\
 \hline
\end{tabular}}
\caption{Properties of the repeat bursts detected from \FRB\ with MeerKAT.\\
a: Label of each burst with "U" indicating detection at UHF and "L" at L-band. The highlighted bursts ($\ast$) were not detected by the real-time search pipeline, but identified in the visual inspection of other bursts.\\
b: Time of arrival in Barycentric Dynamical Time referenced to infinite frequency.\\
c: Reported S/N by the real-time search pipeline.\\
d: Boxcar equivalent burst width.\\
e: Fluence of each burst estimated by the radiometer equation and the associated error (see Section~\ref{sec:fluence}).\\
f: "Y" and "N" indicate whether the burst triggered the voltage buffer dump or not (see Section~\ref{sec:voltage}).\\
g: Linear and circular polarisation fraction measured from the voltage data (see Section~\ref{sec:polarimetry}). Burst L9 did not trigger voltage data, but follows burst L10 so closely ($\sim50$\,ms; see Figure~\ref{fig:bursts}) that we can measure its polarisation properties using the voltage data of L10. The same applies to burst U44, but we do not provide measurements due to its low S/N.}
\label{tab:bursts}
\end{threeparttable}
\end{table*}

\renewcommand{\thetable}{\arabic{table} (Continued.)}
\addtocounter{table}{-1}

\begin{table*}
\centering
\resizebox{1.5\columnwidth}{!}{\hspace{-0cm}\begin{tabular}
{l c r r c c c c c}
\hline
 Burst & TOA & $\text{S/N}_\text{det}$ & Burst width & Fluence & Error & Trigger & $L/I$ & $|V|/I$ \\
  & (MJD)&  & (ms) & (Jy ms) & (Jy ms) &  & & \\
\hline
 L1 & 60349.36537245 & 27.8 & 4.5 & 1.95 & 0.07 & Y & 0.95(9) &0.17(5) \\
 L2 & 60349.36600170 & 63.7 & 2.9 & 1.48 & 0.05 & Y &1.02(8) &0.09(4) \\
 L3 & 60349.36671720 & 77.0 & 4.5 & 4.21 & 0.07 & Y &0.93(4) &0.21(2) \\
 L4 & 60349.36720463 & 39.5 & 4.5 & 1.56 & 0.07 & Y &1.06(10) &0.09(4) \\
 L5 & 60349.36731330 & 23.8 & 3.6 & 1.53 & 0.06 & Y &1.06(21) &0.12(7) \\
 L6 & 60349.36930761 & 50.1 & 4.5 & 2.07 & 0.06 & Y &1.09(9) &0.14(4) \\
 L7 & 60349.37039322 & 28.5 & 17.0 & 2.32 & 0.07 & Y & 1.04(9)&0.11(4) \\
 L8 & 60349.37583829 & 13.2 & 3.6 & 0.61 & 0.06 & Y &0.95(17) &0.16(8) \\
 L9 & 60349.38019587 & 25.5 & 4.5 & 1.05 & 0.06 & N &0.97(12) &0.14(6) \\
 L10 &60349.38019691 & 57.6 & 5.6 & 2.96 & 0.07 & Y &0.94(6) &0.08(3) \\
 L11 &60349.38032034 & 13.2 & 4.5 & 1.13 & 0.13 & Y &0.94(24) &0.08(11) \\
 L12 &60349.38280178 & 16.2 & 5.6 & 1.40 & 0.15 & Y &1.00(19) &0.20(10) \\
 L13 &60349.38816473 & 11.1 & 3.6 & 0.75 & 0.07 & Y &1.14(31) &0.24(13) \\
 L14 &60349.38915758 & 28.7 & 2.3 & 0.80 & 0.05 & Y & 0.92(15)&0.06(6) \\
 L15 &60349.39190003 & 43.6 & 7.0 & 1.97 & 0.13 & Y &1.00(5) &0.16(3) \\
 L16 &60349.39266883 & 14.9 & 3.6 & 1.21 & 0.15 & Y &1.27(27) &0.20(9) \\
 L17 &60349.39467029 & 32.6 & 4.5 & 1.05 & 0.08 & Y &0.96(10) &0.11(5) \\
 L18 &60349.39728630 & 17.1 & 2.3 & 0.83 & 0.12 & Y &1.12(20) &0.12(9) \\
  \hline
\end{tabular}}
\caption{}
\end{table*}

\renewcommand{\thetable}{\arabic{table}}

\renewcommand{\thefigure}{\arabic{figure} (Continued.)}
\addtocounter{figure}{-1}

\begin{figure*}
\centering
\includegraphics[width=.85\textwidth]{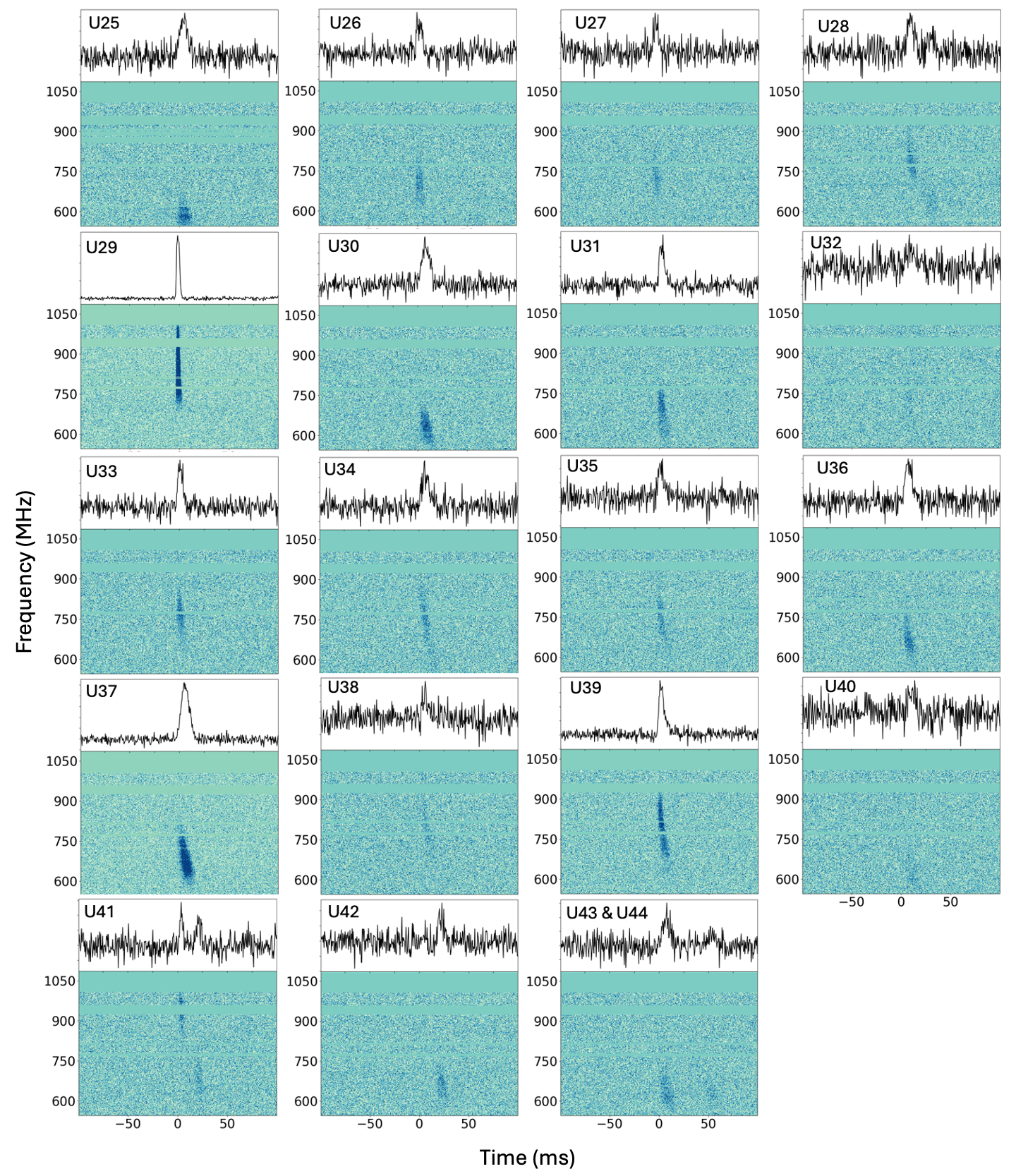}
\caption{}
\end{figure*}

\renewcommand{\thefigure}{\arabic{figure}}

\renewcommand{\thefigure}{\arabic{figure} (Continued.)}
\addtocounter{figure}{-1}

\begin{figure*}
\centering
\includegraphics[width=.85\textwidth]{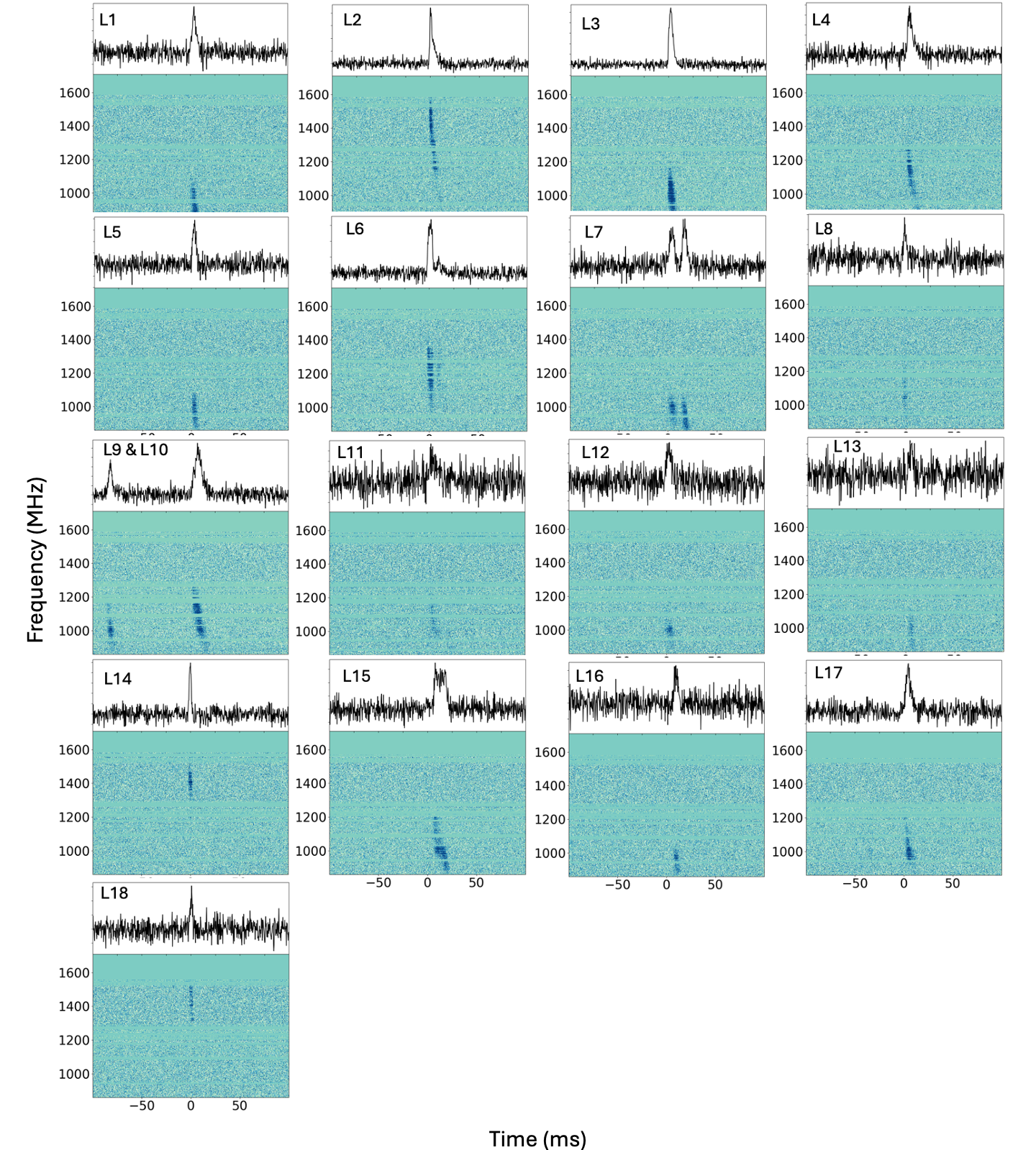}
\caption{}
\end{figure*}

\renewcommand{\thefigure}{\arabic{figure}}

\subsection{Voltage data}\label{sec:voltage}
We triggered on 48 of the 62 detected bursts, including 31 at UHF and 17 at L-band (see Table~\ref{tab:bursts}), and downloaded the voltage data along with the gain solutions for all available frequency channels. In order to localise the FRB source, we correlated the voltage data to create visibilities after applying the gain solutions. Images were made with {\sc wsclean} \citep{Offringa14, Offringa17} around the time of the burst detection with an integration time of 0.96\,ms and 0.61\,ms at UHF and L-band, respectively. Any bright $\sim$ms bursts can be identified as transient sources in these images. Note that we did not perform flux density calibration for the images, but that does not affect our localisation of the FRB source. Further details on the imaging methods can be found in \citealt{Rajwade24}. 

After determining the FRB position through imaging (see below), we coherently beamformed the voltage data at the best FRB coordinates
to create a high time-resolution, full polarization time-frequency data product for each burst. We used the {\sc dspsr} package \citep{Straten11} to write these beamformed data into an archive format that can be processed using tools from {\sc psrchive} \citep{Hotan04}, including {\sc pazi} for removing RFIs and {\sc pam} for transforming to Stokes parameters. Compared to the real-time data products, these archive data have a much higher resolution in time and frequency ($0.13\,\text{MHz}/7.5\,\mu\text{s}$ and $0.21\,\text{MHz}/4.8\,\mu\text{s}$ at UHF and L-band, respectively) and full Stokes information. In addition, as we can beamform right at the interferometric localisation of \FRB\ using the 64 antennas 
(see Section~\ref{sec:imaging}), and perform coherent dedispersion with {\sc dspsr}, we expect the pulse signals to be much stronger in these archive data than detected in the filterbank data. All these facilitate studying the dynamic pulse structure of the bursts with voltage buffer dumps and their polarisation properties. See \citealt{Rajwade24} for more details on the beamforming process.

\section{Results}

\subsection{DM estimation}\label{sec:DM}

Upon detection of the \FRB\ bursts, we estimated the DMs that maximised the peak S/Ns of individual pulses. However, these S/N maximising DMs could result in loss of intrinsic temporal sub-structure, especially for those bursts with complex morphology and multiple components. We therefore used a structure optimising approach to measure the DM of \FRB. We selected the two brightest bursts in our sample, U29 detected in the UHF with a S/N of 177.08 and showing a simple pulse and L3 detected in the L-band with a S/N of 77.02 and showing two sub-components (see Figure~\ref{fig:bursts}), and ran {\sc DM\_phase}\footnote{\url{https://github.com/danielemichilli/DM_phase}} \citep{Seymour19}, a DM optimisation algorithm that maximises the coherent power across the bandwidth. We dedispersed the two bursts over a trial DM range of $520\text{--}535\,\text{pc}\,\text{cm}^{-3}$ in steps of $0.01\,\text{pc}\,\text{cm}^{-3}$, and obtained a structure optimising DM of $527.65\pm0.01\,\text{pc}\,\text{cm}^{-3}$ and $527.92\pm0.14\,\text{pc}\,\text{cm}^{-3}$ for the two bursts, respectively. The uncertainty on each DM was calculated through transforming the standard deviation of the coherent power spectrum into a standard deviation in DM. Given that these two DM values agree within a $2\sigma$ error, we adopt $\text{DM} = 527.65\pm0.01\,\text{pc}\,\text{cm}^{-3}$ for \FRB\ throughout this work. This DM is also consistent with that reported by CHIME/FRB.

\begin{figure}
\centering
\includegraphics[width=.5\textwidth]{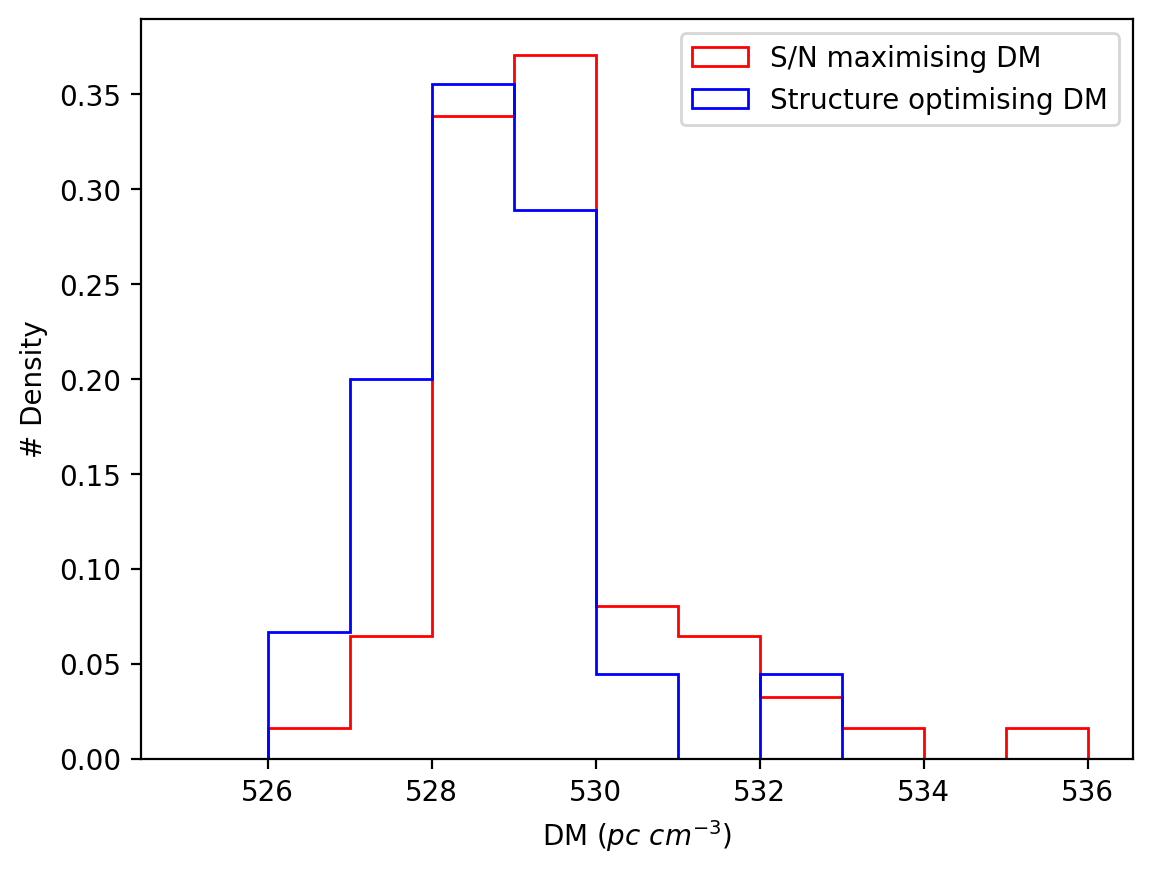}
\caption{Distribution of S/N and structure optimising DMs for the repeat bursts detected from \FRB. This plot shows the DM variation within the sample of bursts (see Section~\ref{sec:DM}).}
\label{fig:DM_distribution}
\end{figure}

For comparison, we computed both the S/N and structure optimising DMs for all the bursts. The distribution is shown in Figure~\ref{fig:DM_distribution}. Note that we do not provide structure optimising DMs for the bursts with a low S/N ($\lesssim13$), as this approach does not produce a good fit of the S/N--DM curve. As can be seen, most of the bursts show a DM in the range of $528\text{--}530\,\text{pc}\,\text{cm}^{-3}$, which is consistent with our observation of many under-dedispersed bursts in Figure~\ref{fig:bursts}.
This DM discrepancy corresponds to a dispersion delay of $\sim10$\,ms across the emission bandwidth, comparable to the pulse width of the bursts. Therefore, we do not expect significant distortion in the burst structure due to the DM variation observed here. Also, the bursts we detected from \FRB\ do not show narrow components and many of them have a single component, so the structure optimising algorithm does not work well on them. Additionally, the observed difference in DM is likely to be an intrinsic property of the bursts, since the DM is not expected to evolve on a timescale of a few minutes. In summary, we conclude that the DM measured from the brightest burst can be applied to all the bursts of \FRB\ analysed here.

\subsection{Imaging and localisation}\label{sec:imaging}

\begin{figure*}
\subfigure[Burst L2]{
  \label{fig:imaging_L2}
  \includegraphics[width=.9\linewidth]{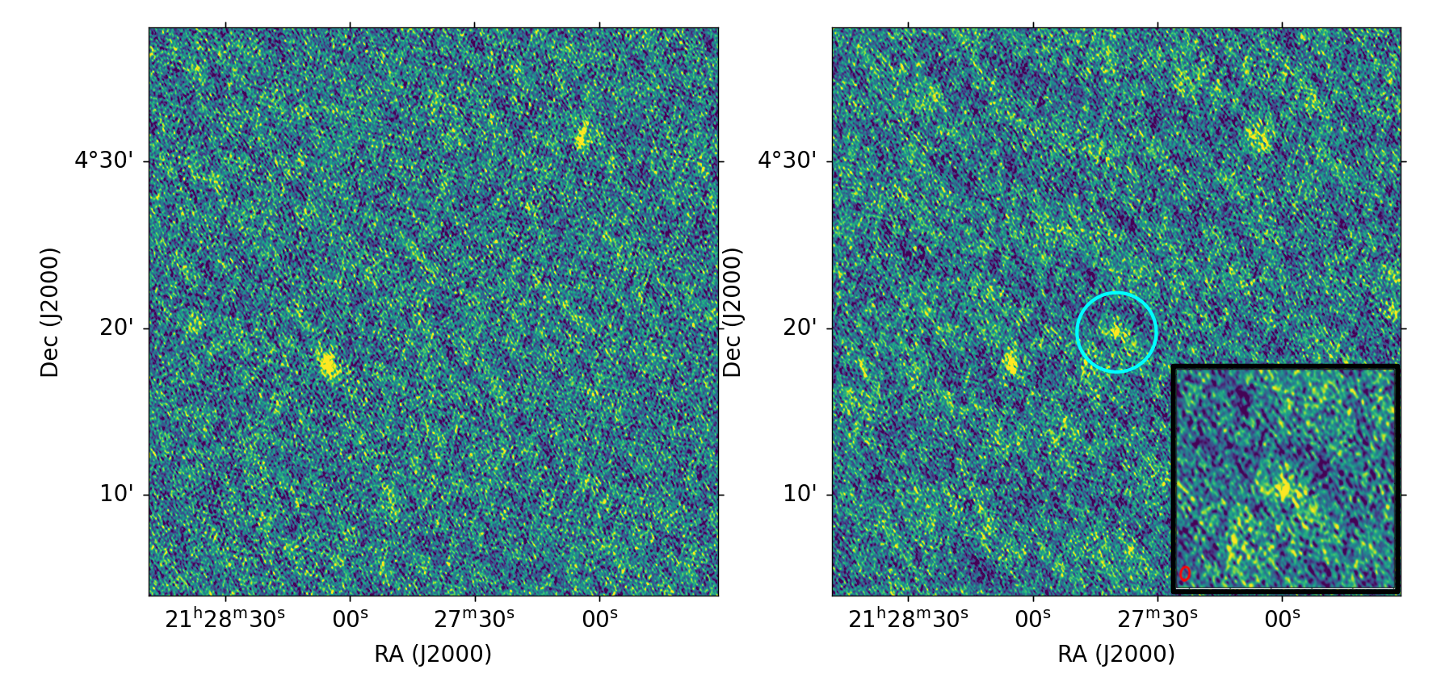}} \\
\subfigure[Burst L3]{
  \label{fig:imaging_L3}
  \includegraphics[width=.9\linewidth]{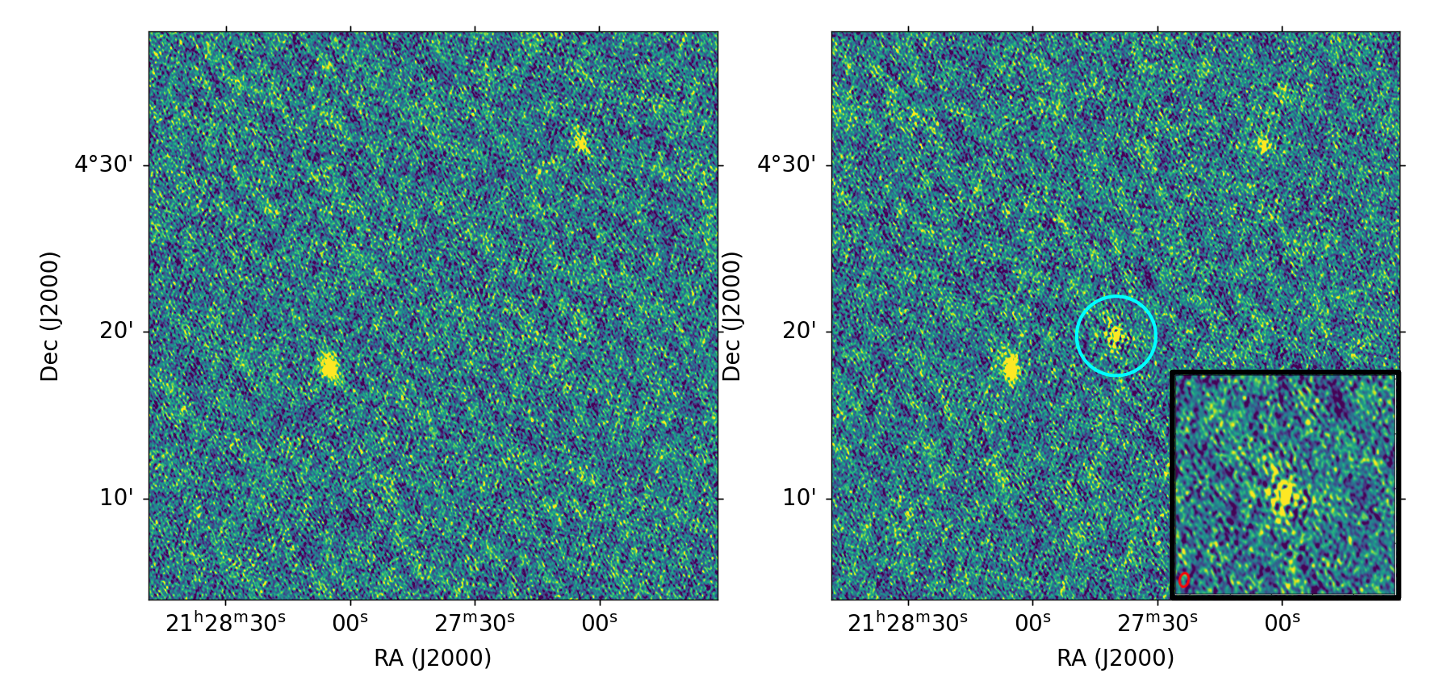}}
\caption{Images of the position of \FRB\ integrated over the duration of burst L2 and L3 (right) and before the burst detection (left). The cyan circle (2\,arcmin radius) marks the transient source identified at the time of the FRB detection. The images have a synthesised beam size of $15\,\text{arcsec}\times10\,\text{arcsec}$. The inset at the bottom right corner of the on-source image is a zoomed in view to display the FRB source. The red ellipse at the bottom left corner of the inset indicates the synthesised beam.}
\label{fig:imaging}
\end{figure*}

We selected the two brightest bursts detected in the L-band, L2 and L3, for imaging and localisation of the \FRB\ source. Note that in the L-band we can achieve a better localisation accuracy than in the UHF given the higher frequencies and thus higher resolution. 
For each burst, we created two images, "on" and "off", with the voltage data covering the same duration of and before the burst respectively, as shown in Figure~\ref{fig:imaging}.
We adopted a pixel scale of 1\,arcsec and size of $8192\times8192$ pixels, and used the {\sc wsclean} algorithm \citep{Offringa14, Offringa17} for deconvolution. This imaging exercise also provides a check on the data quality. We identified a transient source in the on-burst image for both bursts, as indicated by the cyan circle in Figure~\ref{fig:imaging}. Given this source appears only at the time of the FRB detection and at the same position for different bursts, and since there is no other transient source in the images, we confirm that this is the FRB source.

In order to localise the FRB source, we need to perform an astrometric correction on the MeerKAT images. 
We imaged the full 300\,ms of voltage data for each burst. The Python Blob Detector and Source Finder\footnote{\url{https://www.astron.nl/citt/pybdsf/}} ({\sc pybdsf}) algorithm was used to find the positions of all sources in these images, which were then matched to the sources in the Rapid ASKAP Continuum Survey (RACS) catalogue \citep{McConnell20}, excluding any sources that appear resolved in the MeerKAT images. This resulted in 16 matches with the separations ranging from 0.5 to 2.9\,arcsec. We used the matched sources to solve for a transformation matrix to shift and rotate the MeerKAT sources to match the RACS source positions\footnote{The code for performing the astrometric correction can be found on GitHub: \url{https://github.com/AstroLaura/MeerKAT_Source_Matching}}. After the astrometric correction, the separations between the matched sources reduced to $0.3\text{--}2.5$\,arcsec with a median of 1\,arcsec. We then applied the transformation matrix to the $\sim\text{ms}$ images containing burst L2 and L3. Running {\sc pybdsf} on these astrometry corrected images, we found the source position to be RA = 21:27:39.82, Dec = +04:19:45.93 and RA = 21:27:39.86, Dec = +04:19:45.01 for L2 and L3, respectively. The uncertainty on the RA and Dec given by {\sc pybdsf} is $\sim0.1$\,arcsec, much smaller than the absolute astrometric uncertainty from the RACS positions (1\,arcsec in both RA and Dec) and the median offset of the positions after the astrometric correction (1\,arcsec). We added these uncertainties in quadrature and found the total uncertainty to be 1.4\,arcsec. 

The two positions obtained for \FRB\ are consistent with each other within the uncertainty, and $\sim1.2$\,arcmin away from the best CHIME/FRB position (within their $3\sigma$ error region). 
This is consistent with the position we quoted in ATel \#16446 \citep{Tian24} and was later confirmed through the VLBI localisation within the PRECISE project \citep{Snelders24}.
In the following analysis we use the position of the brightest burst we detected in the L-band (L3) for \FRB.

\subsection{Burst morphology}\label{sec:morphology}

The sample of repeat bursts detected from \FRB\ display a wide range of burst morphologies, as shown in Figure~\ref{fig:bursts}. 
Note that there is an ambiguity in distinguishing whether two components are separate bursts or belong to the same burst. In this work, we consider two bursts being independent if they are separated by $\gtrsim50$\,ms (twice the widest pulse in our sample).
While most of the bursts show single-peaked pulse profiles, some comprise multiple components, e.g. bursts U7, U11, U28, U41, L6 and L7. 
While the sub-components in U7, U28 and U41 seem to follow the downward drifting trend usually observed in repeating FRBs \citep{Pleunis21b}, the burst U11 shows upward drifting sub-components. This morphology has previously been observed in FRB 20121102A \citep{Platts21} and FRB 20201124A \citep{Kumar22}, although it is quite rare. 
We also observe a frequency downward drift in most of the single-peaked repeat bursts. Taking the bright burst U37 as an example, we used the 2D autocorrelation function (ACF) to estimate the linear drift rate, yielding a value of $\Delta f/\Delta t=-25.5\,\text{MHz}\,\text{ms}^{-1}$. The uncertainty is not well constrained with the direct Gaussian fitting approach, so is not included here. Other bursts show a drift rate between $\sim-0.1$ and $\sim-34\,\text{MHz}\,\text{ms}^{-1}$. 

The observed downward drifting in the bursts of FRB 20240114A can be simply explained using radius-to-frequency mapping \citep{Wang19}. In this model, the FRB emission is produced by charged particles in the magnetosphere of a neutron star (NS) through curvature radiation. As the charged particles move away from the NS, the curvature radius continuously increases, corresponding to a decrease in the curvature radiation frequency. Assuming several bunches of charged particles are launched simultaneously along adjacent field lines from the NS surface, our line of sight will always catch the lower-altitude (i.e., higher frequency) emission first and the higher-latitude (i.e., lower frequency) emission later, resulting in the observed downward drifting. However, occasional upward drifting could be observed in the case that the bunches are launched at different times \citep{Wang20b}. This is consistent with the observed downward drifting and occasional upward drifting in the bursts of \FRB, and might suggest curvature radiation of electron-positron pairs within the magnetosphere of a NS to be the FRB emission mechanism.

\begin{figure*}
\centering
\includegraphics[width=\textwidth]{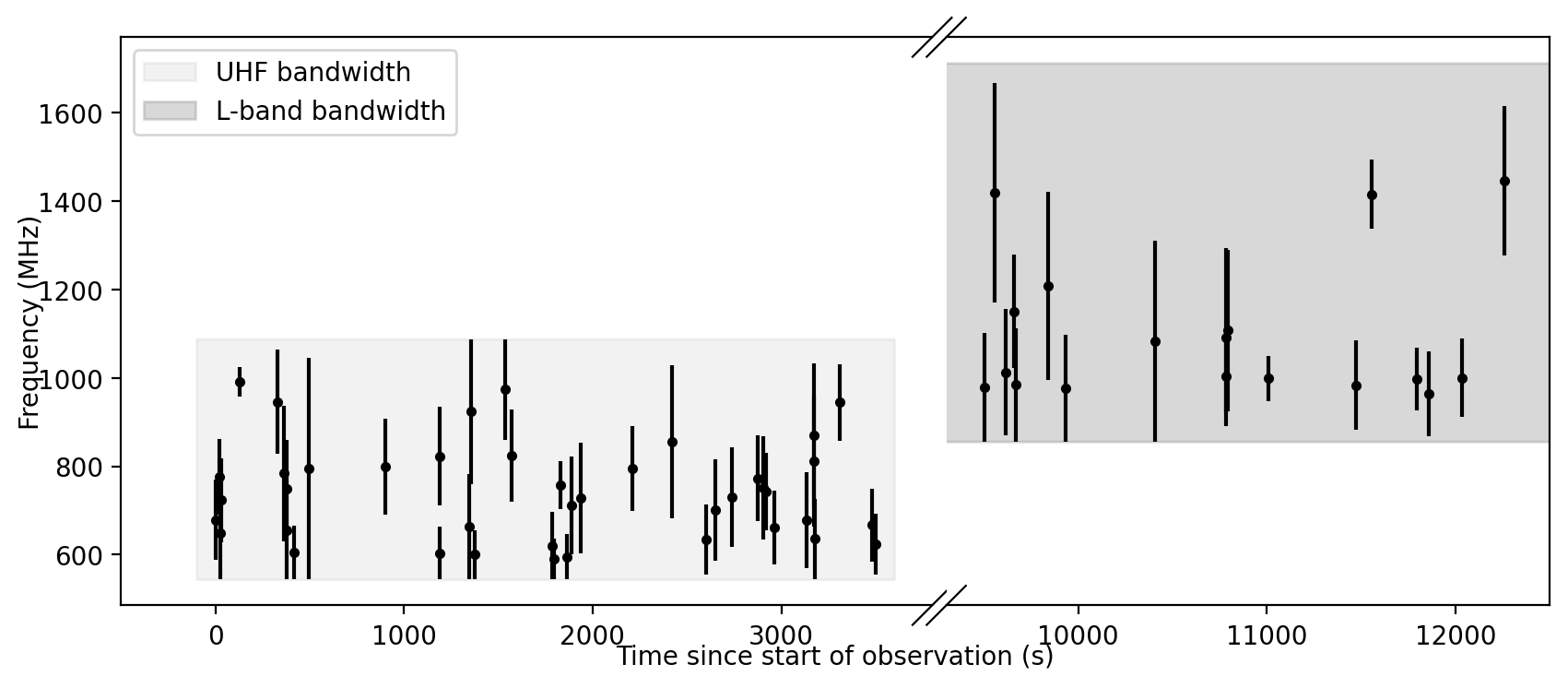}
\caption{Burst spectral extent as a function of burst arrival time. Each point represents the central emission frequency of a burst with the error bar corresponding to the spectral extent. We did not measure the spectral extent of the three low-S/N events, U12, U19 and U44. The UHF bursts are separated from the L-band bursts by a $\sim100$\,min gap. The light and dark shaded regions indicate the bandwidth of the UHF and L-band observations, respectively.}
\label{fig:spec_extent}
\end{figure*}

A critical distinction between repeating FRBs and apparent one-off events is their spectral extent, with repeaters being usually narrower in frequency than one-offs \citep{Pleunis21b}.
Figure~\ref{fig:spec_extent} shows the spectral extent of the repeat bursts detected from \FRB\ as a function of burst arrival time. Note that there is a $\sim100$\,min gap between the UHF and L-band observations. We used a Gaussian function to fit the spectrum of the on-pulse region of each burst and adopted the full width at 10\% of the maximum as the spectral extent. We did not fit the spectra of Bursts U12, U19 and U44 due to their low S/Ns.
In comparison to the full bandwidth of the UHF (544\,MHz) and L-band (856\,MHz), all the repeat bursts detected from \FRB\ are band limited, with the average spectral occupancy being $\sim210$\,MHz at UHF and $\sim270$\,MHz at L-band, corresponding to a fractional bandwidth of $\sim40\%$ and $\sim30\%$, respectively. Note that the signal spectral envelope extends to the top or bottom of the bandwidth in 32 bursts (e.g. U17, U29, L1 and L3), for which our spectral extent measurements should be considered as lower limits.

Repeating FRBs are observed to display a correlation between their emission activity and frequency \citep{Aggarwal20}. Our monitoring of the \FRB\ source reveals a dearth of emission above $\sim1.4$\,GHz, with the majority of bursts centering around $\sim700$\,MHz in the UHF and $\sim1$\,GHz in the L-band, as shown in Figure~\ref{fig:spec_extent}. It is therefore possible that the FRB emission is very faint at higher frequencies. Further sensitive observations are needed to probe the high-frequency activity of \FRB. The preferred emission frequency of some repeating FRBs has been found to evolve on a timescale of hours to days \citep{Gourdji19, Pastor21, Kumar22}. We do not find such a trend within our observation of \FRB.

The repeat bursts from \FRB\ show a wide range of burst widths from 1.84\,ms to 21.4\,ms, as listed in Table~\ref{tab:bursts}. However, the large widths of burst U18 and L7 arise from multiple distinct components being counted as a single burst. Removing these two bursts, we obtained a mean width of 5.92\,ms, comparable to the burst widths of other known repeaters \citep{Pleunis21b}. Also we found the bursts in the L-band seem to be narrower than in the UHF, which may be caused by the intra-channel smearing, $\sim2$\,ms and $\sim4$\,ms at the center of L-band and UHF, respectively.


We do not find strong evidence in the morphology of the single-peaked repeat bursts from \FRB\ for temporal scattering. The scattering timescale along the line of sight of the FRB source, as predicted by the galactic distance models NE2001 and YMW16, is $0.2\,\mu$s and $0.5\,\mu$s, respectively, at 1\,GHz \citep{Cordes02, YMW16, Price21}. We selected burst U29, the brightest burst in our sample and with a single-peaked narrow pulse, to constrain the scattering timescale. We performed scattering fits to this burst using a custom PYTHON-based software {\sc scatfit}\footnote{\url{https://github.com/fjankowsk/scatfit/}}, which convolves a single-sided exponential decay function with a Gaussian base model to create a broadened pulse profile \citep{Jankowski23}, and obtained a scattering timescale of $0.4\pm0.2$\,ms at 1\,GHz, much smaller than the burst width. 
Note that the bandwidth was split into 4 sub-bands, 3 of which have enough S/N to measure the scattering, and the scattering index was fixed to $-4$.
Given that scattering is difficult to model in bursts with multiple components or frequency drifting, we do not attempt to measure scattering timescales for the full sample of bursts detected from \FRB.

The dynamic spectra of several repeat bursts in our sample show scintillation, e.g. U29, U39, L3 and L6. We used the brightest burst U29 to measure the scintillation bandwidth ($\text{BW}_\text{sc}$). First we split the burst spectral extent into 7 sub-bands, from 729\,MHz to 966\,MHz. Then we computed the ACF of the spectral intensity for each sub-band and used a Lorentzian to fit the central peak where the half width at half maximum is the scintillation bandwidth. We obtained $\text{BW}_\text{sc}=4.36\pm0.11$\,MHz at the central frequency of 847\,MHz of the burst emission, and a power-law index of $\gamma=2.84\pm1.45$ for the evolution of $\text{BW}_\text{sc}$ with frequency ($\propto\nu^\gamma$), which is consistent with the expected value of $\gamma=4$ for scattering in a turbulent plasma. 
Along the line of sight of \FRB, the NE2001 and YMW16 galactic models predict a scintillation bandwidth of $\sim0.5$\,MHz at 1\,GHz \citep{Cordes02, YMW16}. This suggests that the observed spectral modulation might arise from interstellar scintillation in the FRB host rather than the Milky Way. However, it is worth noting that the Galactic electron density models have large uncertainties at high Galactic latitudes \citep{Gaensler08}, which is the case for FRB 20240114A ($b\approx-31.7$\,deg). With a frequency resolution of 0.13\,MHz at UHF that is comparable to the predicted decorrelation bandwidth from the Milky Way interstellar medium, we cannot rule out the models completely.
\subsection{Fluences}\label{sec:fluence}

We estimated the fluence of each burst using the modified single-pulse radiometer equation \citep{Dewey85} from \cite{Jankowski23}:

\begin{equation}
    S_\text{peak} (\text{S/N}, W_\text{eq}, \vec{a}) = \text{S/N}\,\beta\,\eta_\text{b}\,\frac{\text{SEFD}}{\sqrt{b_\text{eff}N_\text{p}W_\text{eq}}}\,a^{-1}_\text{CB}\,a^{-1}_\text{IB},
    \label{eq:radiometer}
\end{equation}

\noindent where $S_\text{peak}$ is the peak flux density, $\vec{a}=(a_\text{CB}, a_\text{IB})$ are the attenuation factors of the detection CB and incoherent beam (IB), $\beta\approx1$ is the digitization loss factor, $\eta_\text{b}\approx1$ is the beamforming efficiency, $\text{SEFD}$ is the system equivalent flux density of the MeerKAT array\footnote{See the online MeerKAT technical documentation: \url{https://science.ska.ac.za/meerkat}}, $b_\text{eff}$ is the effective bandwidth in Hz, $N_\text{p}=2$ is the number of polarisations, and $W_\text{eq}$ is the observed equivalent boxcar pulse width in seconds. In the equation above, three parameters are frequency dependent: $\text{SEFD}$, $a_\text{CB}$ and $a_\text{IB}$. 
In order to obtain the frequency-dependent SEFD, we used the polynomial models from \citet{Geyer21}, which for the L-band is
\begin{equation}
    \text{SEFD} = 5.71\times10^{-7}\,\nu^3 - 1.90\times10^{-3}\,\nu^2 + 1.90\,\nu -113,
\end{equation}

\noindent and for the UHF (private communication) is:

\begin{equation}
    \text{SEFD} = 2.30\times10^{-6}\,\nu^3 - 4.69\times10^{-3}\,\nu^2 + 2.52\,\nu + 286,
\end{equation}

\noindent where $\nu$ is the observing frequency in MHz and $\text{SEFD}$ in Jy. Note that the above $\text{SEFD}$ is for a single MeerKAT dish, and we need to divide that by the number of antennas used for the observation (40; see Section~\ref{sec:observation}).
We used the primary beam model ({\sc katbeam}\footnote{\url{https://github.com/ska-sa/katbeam}}) and the coherent beam model ({\sc mosaic}\footnote{\url{https://github.com/wchenastro/Mosaic}}; \citealt{Chen21}) of MeerKAT to calculate the beam correction $a_\text{IB}$ and $a_\text{CB}$ at the position of \FRB\ obtained in Section~\ref{sec:imaging}.

In order to take into account the limited spectral extents of the repeat bursts of \FRB, we split the full bandwidth into 8 sub-bands and estimated the fluence in each sub-band separately before adding them up. Specifically, we measured the S/N of the burst dedispersed to the optimal DM (see Section~\ref{sec:DM}) in each sub-band and converted it to a flux density using Eq.~\ref{eq:radiometer} and fluence $F=S_\text{peak}\,W_\text{eq}$. Note that $\text{SEFD}$, $a_\text{CB}$ and $a_\text{IB}$ have different values in different sub-bands, and $b_\text{eff}$ is the effective bandwidth of the sub-band excluding flagged channels. The final fluence of each burst summed from all the sub-bands is listed in Table~\ref{tab:bursts}, with the uncertainty corresponding to $1\sigma$ radiometer noise.

\begin{figure}
\centering
\includegraphics[width=.5\textwidth]{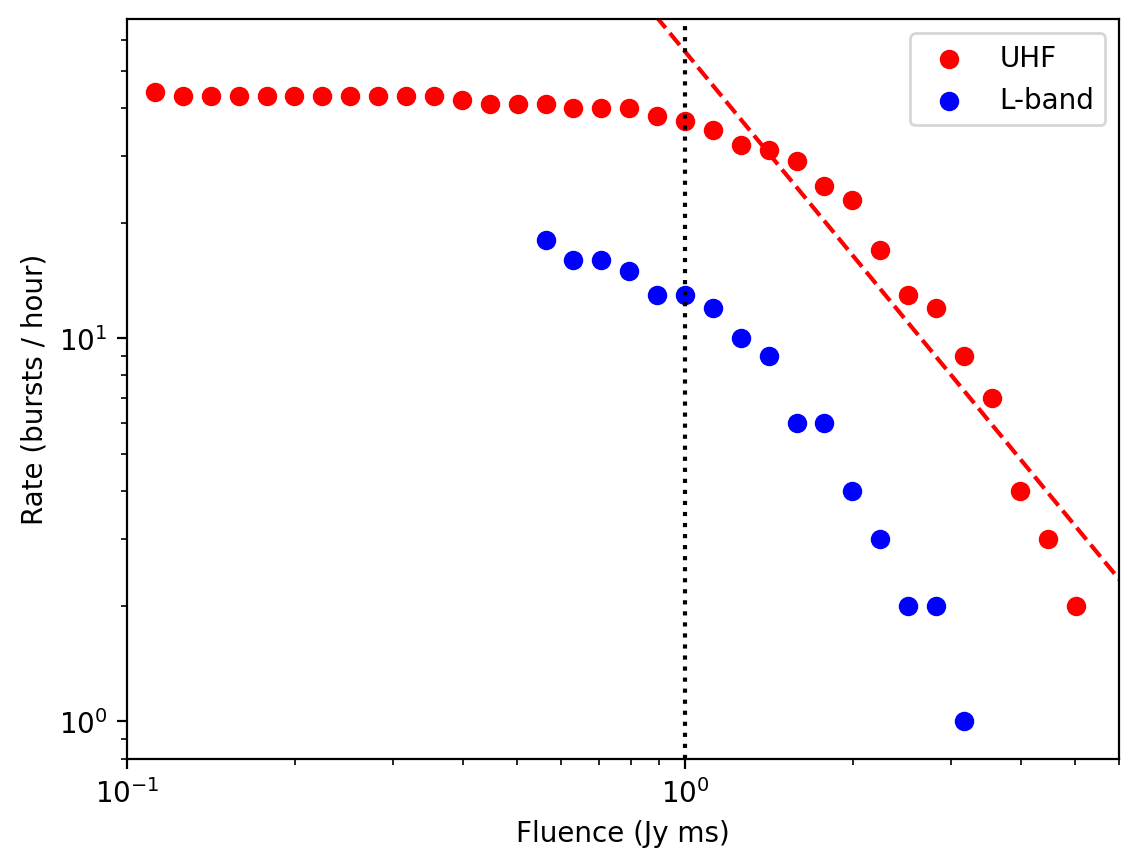}
\caption{Cumulative burst rate function of the MeerKAT detected bursts at UHF and L-band. The vertical dotted line marks the fluence completeness limit of MeerKAT observations, and the dashed red line shows the best-fitting power law for the bursts above the completeness level at UHF. No fit is provided for the small sample of L-band bursts.}
\label{fig:fluence}
\end{figure}

Figure~\ref{fig:fluence} shows the cumulative distribution of the MeerKAT detected burst rate at UHF and L-band above a given fluence. There are two power-law features in the distribution, with the break being visually identified at $\sim1$\,Jy\,ms. We attribute this break to the completeness limit of MeerKAT observations and model the cumulative distribution with a function of the form $R (>F)\propto F^\gamma$. There is a detailed study of the MeerTRAP survey performance and completeness, where the fluence completeness limit is estimated to be 0.7\,Jy\,ms \citep{Jankowski23}, which is consistent with the value observed here. Our fit excluding bursts below the completeness limit gives a power-law index of $\gamma=-1.8\pm0.2$ at UHF. We do not provide fitting results for the small sample of L-band bursts. The power-law index determined here is comparable to that measured for FRB 20121102A at 1.4\,GHz ($-1.8\pm0.3$; \citealt{Gourdji19, Aggarwal21}) and FRB20180916B at 600\,MHz ($-1.3\pm0.3\pm0.1$; \citealt{CHIME20b}), but lower than the $\gamma=-1.4\pm0.1$ index found in \citet{Pastor21} for FRB 20180916B. Note that setting a different break in fluence would affect the fitting result, e.g., a break at 2\,Jy\,ms would give a power-law index of $\gamma=-2.6\pm0.2$, steeper than that with the 1\,Jy\,ms break. In that case, it might suggest a turnover in the burst energy distribution, similar to that observed for FRB 20121102A \citep{Li21}. A more rigorous analysis of fluence completeness is beyond the scope of this paper.

\subsection{Burst rate and arrival times}

We detected 44 and 18 bursts in the 1\,hr MeerKAT UHF and L-band observation, respectively, with their TOAs being listed in Table~\ref{tab:bursts}. If all bursts above our detection limit, 0.12\,Jy\,ms at UHF and 0.61\,Jy\,ms at L-band (the faintest bursts detected in these two bands), have been detected, we obtain a burst rate of $44\,\text{hr}^{-1}$ and $18\,\text{hr}^{-1}$ at UHF and L-band, respectively. However, these estimates are likely to be inaccurate due to the incomplete fraction $\lesssim1$\,Jy\,ms at UHF (see Section~\ref{sec:fluence}) and the small sample at L-band. Considering only the bursts above the completeness limit
of $\sim1$\,Jy\,ms at UHF, we arrive at a burst rate of $\sim37\,\text{hr}^{-1}$.
For comparison, FAST observations of \FRB\ between 1--1.5\,GHz report an average burst rate of $\sim20\,\text{hr}^{-1}$ above the 0.015\,Jy\,ms fluence threshold between 2024 January 28 and February 4 \citep{Zhang24} and $\sim500\,\text{hr}^{-1}$ on 2024 March 5 \citep{Zhang24b}, suggesting the FRB activity to be extremely variable. uGMRT observations between 550--700\,MHz report a burst rate of $\sim8\,\text{hr}^{-1}$ above a fluence of 0.6\,Jy\,ms on 2024 February 25 \citep{Panda24}, comparable to our measurement.
All these quoted burst rates qualify \FRB\ as a hyperactive repeater, similar to FRB 20121102A with a peak burst rate of $122\,\text{hr}^{-1}$ observed by FAST above $0.06$\,Jy\,ms \citep{Li21} and FRB 20201124A with a rate of $16\,\text{hr}^{-1}$ by uGMRT above $7$\,Jy\,ms \citep{Marthi21}.
We also notice the \FRB\ source is much more active in the UHF than in the L-band, as can be seen in Figure~\ref{fig:fluence}. Assuming the same fluence limit for the L-band as in the UHF, we find \FRB\ emits $3\times$ more bursts at 816\,MHz than at 1284\,MHz. This frequency-dependent activity has also been observed in other repeaters, e.g. FRB 20180916B, where the FRB source emits over $10\times$ more bursts of the same fluence at 150\,MHz than at 1.4\,GHz \citep{Pastor21}, and has been used to test the radius-to-frequency mapping model in the context of long-period magnetars \citep{Bilous24}.

\begin{figure}
\centering
\includegraphics[width=.5\textwidth]{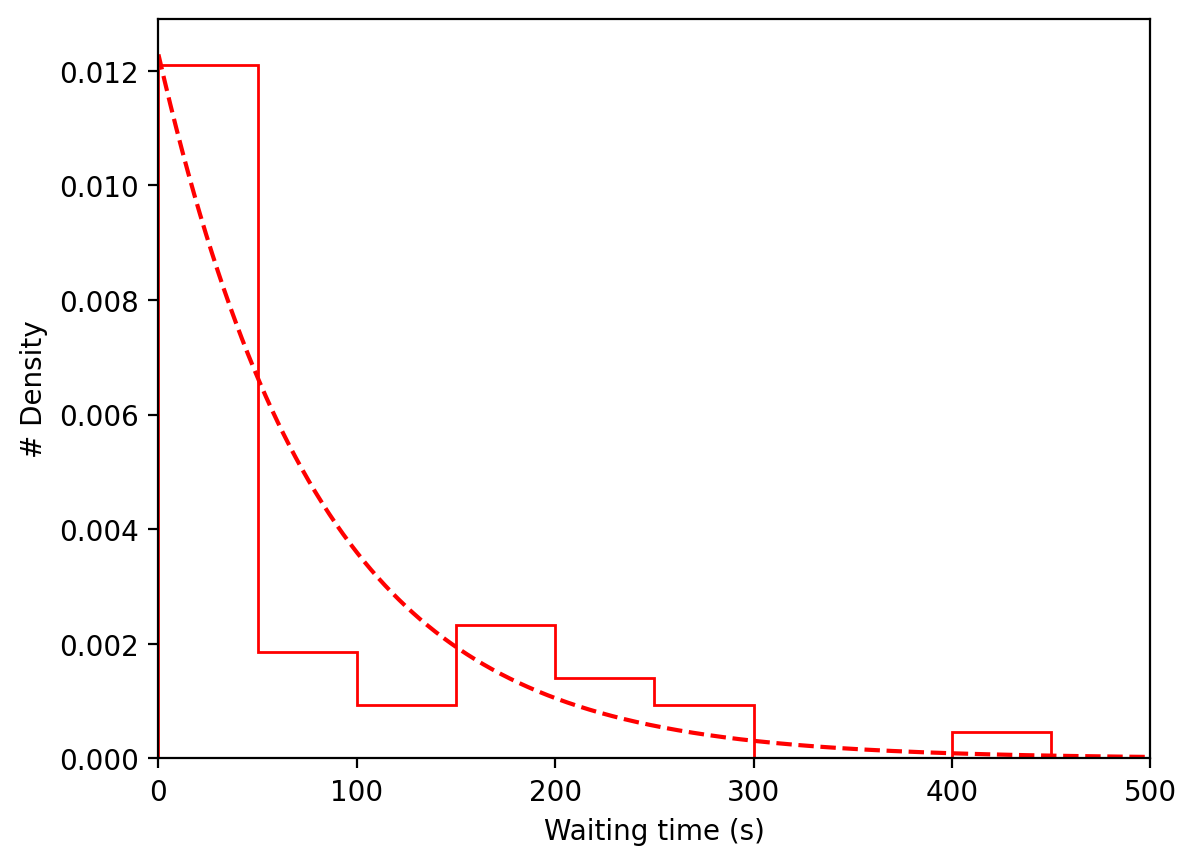}
\caption{Distribution of the burst waiting time in the MeerKAT UHF observation. The dashed line shows the Poisson distribution with a constant rate given by the mean waiting time $\lambda\sim1/81\,\text{s}^{-1}$.}
\label{fig:wait_distribution}
\end{figure}

Figure~\ref{fig:wait_distribution} shows the distribution of waiting times between adjacent bursts observed in the MeerKAT UHF observation. Note that we consider two bursts being separate only if their separation is greater than 50\,ms (see Section~\ref{sec:morphology}). We do not show the distribution for the L-band bursts due to their smaller sample. The mean waiting time between bursts within the 1\,hr UHF observation is $\sim81$\,sec. If the burst occurrence follows a Poisson process, the waiting time should be exponentially distributed with a probability density function 

\begin{equation}
f(t)=\lambda\exp{(-\lambda t)},
\label{eq:poisson}
\end{equation}

\noindent where $\lambda$ is a rate parameter. With only 44 bursts, of which only 37 are above the completeness limit, we do not attempt to model the burst waiting time distribution. We limit our discussion to a comparison between the observed distribution and the Poisson distribution with a constant rate given by the mean value of the waiting times between successive bursts (see above) $\lambda\sim1/81\,\text{s}^{-1}$, as shown in Figure~\ref{fig:wait_distribution}. 
We found weak evidence for the observed burst waiting times being consistent with the constant-rate Poissonian repetition.
This is supported by the two-sided Kolmogorov–Smirnov (KS) test that measures the maximum distance between the empirical cumulative distribution function (CDF) and the CDF corresponding to Eq.~\ref{eq:poisson}, which yields a $p$-value of $\approx0.07$. A smaller $p$-value closer to zero means the two samples are more likely to arise from different distributions. Note that this result only applies to the 1\,hr window of the MeerKAT observation during which \FRB\ is extremely active. Long-term monitoring of the FRB source may reveal any clustering behavior in the burst repetition during bursting activity, as has been demonstrated for other repeaters (e.g. \citealt{Oppermann18, Lanman22}).

We also performed a periodicity search to look for any periodic pattern between the TOAs of the bursts at 1284~MHz and 816~MHz. First, the ToAs were corrected to the solar system barycentre. Then, we made use of the \textsc{altris} bruteforce period fitting algorithm (Morello et al. in prep). \textsc{altris} attempts to fit the integer number of rotations within every gap between consecutive barycentred TOAs, from the shortest to the longest gap. The algorithm attempts to recursively discover the rotation counts; assuming a tentative solution for the k shortest gaps is known, a phase model is fitted to the TOA gaps, from which a range of possibilities for the rotation count k + 1 is calculated. If no integer rotation counts are possible, the tentative solution is discarded; otherwise, the algorithm attempts to further expand the set of new tentative solutions for the first k + 1 gaps. This amounts to a depth-first search of a tree of potential solutions. \textsc{altris} did not find any solution for periods ranging from 50\,ms (the smallest separation between adjacent bursts given our criterion for distinguishing separate bursts; see Section~\ref{sec:morphology}) to the total duration of the MeerKAT observations.

\subsection{Polarimetry}\label{sec:polarimetry}

While the bursts of \FRB\ were detected only in total intensity, the triggered voltage data captured complete polarisation information for some bursts (see Table~\ref{tab:bursts}). 
In order to study the polarisation properties of these bursts, we need to correct for the Faraday rotation by de-rotating the phase in Stokes Q and U produced by the RM in order to measure the polarisation fraction and polarisation position angle (PPA) as a function of time.
We selected the same two bursts as in Section~\ref{sec:DM} and used the {\sc rmsynth} tool from the software {\sc psrsalsa}\footnote{\url{https://github.com/weltevrede/psrsalsa}}, a suite of algorithms for statistical analysis of pulsar data \citep{Weltevrede16}, 
to measure the RM of \FRB. This method transforms polarised intensity as a function of $\lambda^{2}$ to Faraday depth, $\phi$, representing polarised intensity for different trial RMs. We searched for a range of trial RMs between $\pm10000\,\text{rad}\,\text{m}^{-2}$ with a step size of $0.1\,\text{rad}\,\text{m}^{-2}$. Both bursts show a peak in the polarised intensity, corresponding to an RM of $338.1\pm0.1\,\text{rad}\,\text{m}^{-2}$ and $340.5\pm0.5\,\text{rad}\,\text{m}^{-2}$, respectively. The uncertainty on each RM was estimated by the full width at half maximum of the peak in the Faraday depth space. 
These results align with the preliminary analysis reported by CHIME/FRB ($\sim325\,\text{rad}\,\text{m}^{-2}$; \citealt{Shin24}) and Parkes ($\sim360\,\text{rad}\,\text{m}^{-2}$; \citealt{Uttarkar24}).
Similar to the exercise of a global DM for the FRB source in Section~\ref{sec:DM}, here we adopt the RM value measured from the brightest burst in our sample, $338.1\pm0.1\,\text{rad}\,\text{m}^{-2}$, to correct the Stokes spectra of individual bursts and measure their polarisation strength.

We used {\sc psrsalsa} 
to analyse the polarisation data that had been derotated and averaged over frequency (see Section~\ref{sec:voltage}). First we removed the baseline of the Stokes parameters using {\sc pmod}. Then we used {\sc ppol} to calculate the PPA and linear polarisation intensity $L=\sqrt{Q^2+U^2}$. Note that the bias in $L$ for each time sample was removed using \citep{Wardle74}

\[
 L_\text{de-bias} = 
  \begin{cases} 
   L\sqrt{1-\left(\frac{\sigma}{L}\right)^2} & \text{if } 
   L>\sigma \\
   0 & \text{otherwise,}
  \end{cases}
\]

\noindent where $\sigma=\sqrt{(\sigma_Q^2+\sigma_U^2)/2}$ is the off-pulse standard deviation. We set a $3\sigma$ limit on the de-biased $L$ to obtain the significant measurements of PPA. The polarimetric pulse profiles of the 6 brightest bursts in our sample (2 at UHF and 4 at L-band), along with the significant measurements of PPA, are plotted in Figure~\ref{fig:pol_6}. For a full presentation of the polarisation profiles see Appendix~\ref{appendix:pol}. We measured the linear and circular polarisation fractions by averaging $L/I$ and $|V|/I$ across the pulse profile for each burst, as shown in Table~\ref{tab:bursts}. Uncertainties are computed from the off-pulse standard deviation in the Stokes parameters based on the principle of error propagation. Note that we expect polarisation leakages to be small in our measurements given the \FRB\ source is located close to the center of the primary beam ($\sim1.2$\,arcmin; see Section~\ref{sec:imaging}).

\begin{figure*}
\centering
\includegraphics[width=\textwidth]{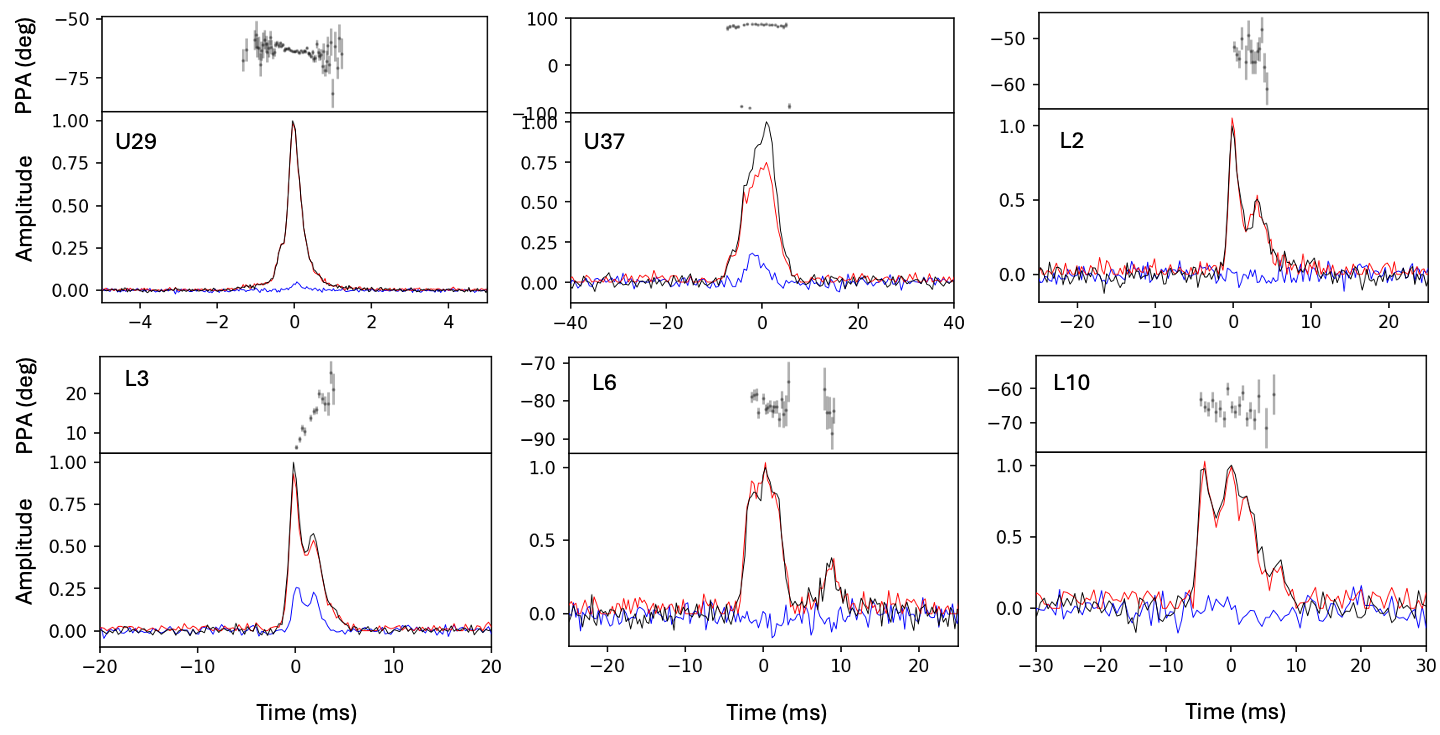}
\caption{Polarimetric pulse profiles of a selection of bursts detected from the \FRB\ source. In each panel, the top shows the PPA, and the bottom shows the frequency averaged pulse profile for total intensity ($I$, black), linear polarisation ($L$, red) and circular polarisation ($V$, blue). The polarisation data are Faraday corrected to the RM value determined in Section~\ref{sec:polarimetry}.}
\label{fig:pol_6}
\end{figure*}

\begin{figure*}
\centering
\includegraphics[width=\textwidth]{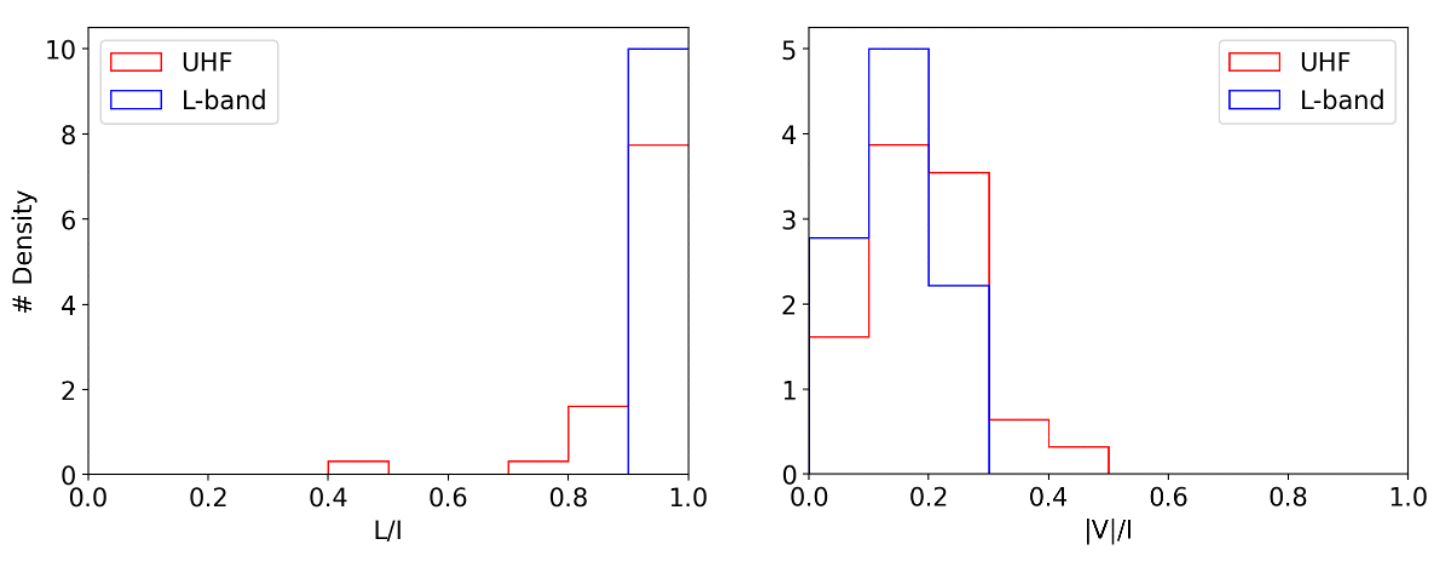}
\caption{Distribution of linear (left) and absolute value circular (right) polarisation of the MeerKAT-detected repeat bursts from \FRB.}
\label{fig:pol_fraction}
\end{figure*}

Figure~\ref{fig:pol_6} shows a diversity of PPA variations across the pulse profiles, e.g. being flat for most of the bursts, sweeping up for burst L3 and sweeping down for burst U29. This is also the case for the other bursts (see Figure~\ref{fig:pol_all}) and reminiscent of the PPA variations observed in bursts detected from FRB 20180301A \citep{Luo20}. 
Figure~\ref{fig:pol_fraction} shows the distribution of $L/I$ and $|V|/I$ for the MeerKAT-detected repeat bursts from \FRB.
We find most of the bursts have $\sim100\%$ degree of linear polarisation and mostly up to $\sim20\%$ degree of circular polarisation, which is consistent with the polarisation fraction measured for bursts detected from this source with FAST \citep{Zhang24}. Whereas highly frequency dependent depolarisation has been observed in eight repeating FRBs \citep{Feng22, Mckinven23}, here we do not find such evidence in the limited sample of repeat bursts detected from \FRB\ as the average $L/I$ in the UHF (0.97) and L-band (1.00) are consistent within the errors. 
We do not find evidence for Faraday conversion either as the average $|V|/I$ in the UHF (0.18) and L-band (0.14) are also consistent within the errors.

\section{Discussion}

\subsection{Polarisation}
\FRB\ shows a high degree ($\sim100\%$) of linear polarisation in all MeerKAT-detected bursts, consistent with the published sample of FRBs \citep{Cordes19, CHIME19b, Fonseca20, Pandhi24}. 
The circular polarisation is low but clearly non-zero in many bursts, with a fraction up to $41\%$ in Burst U11. This is unsurprising given the wide range of circular polarisation fractions up to $\sim70\%$ observed in other active repeaters, e.g. FRBs 20201124A, 20121102A and 20190520B \citep{Hilmarsson21, Feng22b, Jiang22, Kumar22, Xu22}, and suggests circular polarisation could be an essential feature of repeating FRB emission.
We also found a diversity of PPA variations across the pulse profiles of \FRB, with the largest variation spanning a range of $\sim50$\,deg (e.g. U34, U41 and L16; see Figure~\ref{fig:pol_all}). Such a large variation in PPA indicates a varying orientation of the magnetic field with respect to the line of sight, and resembles those seen from radio pulsars and magnetars, implying a magnetospheric origin for the FRB emission \citep{Luo20, Kumar22}.

Depolarisation towards lower frequency has been observed in several repeating FRBs and considered to be caused by multipath propagation in the inhomogeneous magneto-ionic environment \citep{Feng22}. We do not observe this phenomenon in \FRB, which might suggest a different environment. Given the depolarisation has been observed to occur at different frequencies for different FRBs from $<200$\,MHz to $>1$\,GHz, it is possible the depolarisation frequency of \FRB\ is below our observing band. In that case, we would expect a small RM scatter based on the relation between the depolarisation and RM scattering \citep{Sullivan12}, and therefore a less turbulent, dense and/or magnetised environment for \FRB. Low-frequency observations would provide better constraints on the environment of \FRB.

\subsection{Host galaxy}\label{sec:host}

\begin{figure*}
\centering
\includegraphics[width=\textwidth]{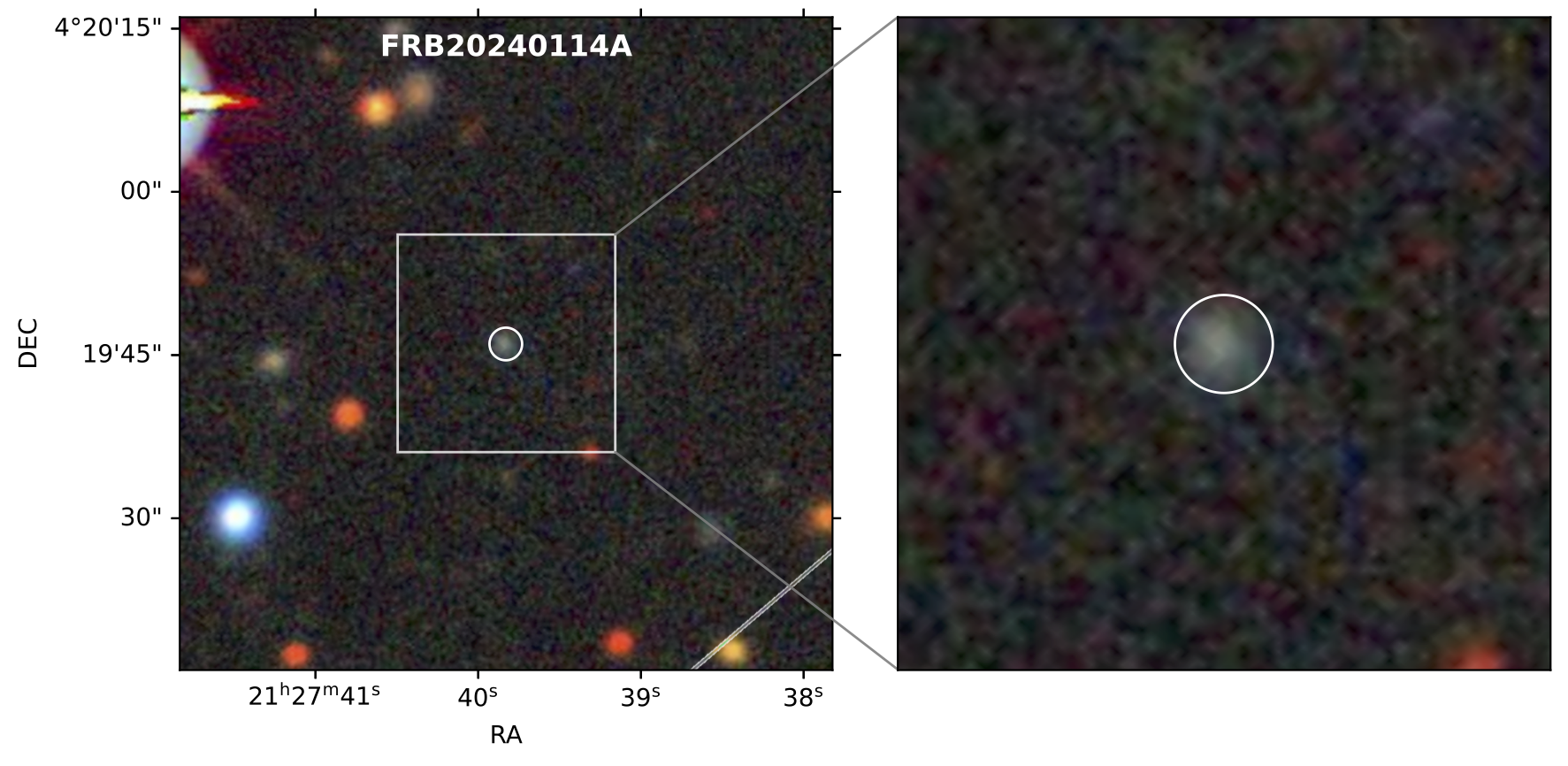}
\caption{DESI Legacy Survey DR10 optical image in the 'grz' filters showing the position of the \FRB\ source (left) and a zoomed in view to display the association of the source with the SDSS galaxy J212739.84+041945.8 (right). The radius of the white circle reflects the uncertainty on the FRB source position.}
\label{fig:optical}
\end{figure*}

We can identify the host galaxy of \FRB\ with the arcsec localisation obtained in Section~\ref{sec:imaging}. Figure~\ref{fig:optical} shows the FRB source on top of an optical image from the DESI Legacy Survey DR10 \citep{Dey19}, with the size of the white circle reflecting the uncertainty on the source position. This allows us to confidently associate \FRB\ with J212739.84+041945.8, a galaxy cataloged in the Sloan Digital Sky Survey (SDSS; \citealt{Alam15}) and with an absolute r-band magnitude of $-17.46\pm0.01$\,AB \citep{Bhardwaj24}. Using the Probabilistic Association of Transients to their Hosts (PATH; \citealt{Aggarwal21b}) software, we found the association probability with this galaxy to be 0.997, assuming the prior on an unseen host of 0.2. This association was later independently confirmed by the EVN-PRECISE team, who localised the FRB source to the same SDSS galaxy with a conservative $\sim200$\,milliarcsecond precision \citep{Snelders24}. We therefore assign
J212739.84+041945.8 as the host of \FRB.

Follow-up observations of the FRB host galaxy with the Optical System for Imaging and low-Intermediate-Resolution Integrated Spectroscopy (OSIRIS) spectrograph at the Gran Telescopio Canarias (GTC) telescope detected various emission lines, leading to a redshift measurement of $z=0.1300\pm0.0002$ \citep{Bhardwaj24}, consistent with the photometric redshift of $z=0.269\pm0.139$ reported in the DESI Legacy survey \citep{Duncan22}. In addition, the FRB host may be a dwarf star-forming galaxy \citep{Bhardwaj24}, similar to the hosts of two active repeating sources, FRB 20121102A and FRB 20190520B \citep{Chatterjee17, Niu22}, which might hint at a potential persistent radio source (PRS) associated with \FRB. Note that \FRB\ has a much lower absolute RM than FRB 20121102A ($1.3\times10^{5}\,\text{rad}\,\text{m}^{-2}$; \citealt{Michilli18}) and FRB 20190520B ($2.4\times10^{4}\,\text{rad}\,\text{m}^{-2}$; \citealt{Anna-Thomas23}), and might be embedded in a less magnetised environment (see Section~\ref{sec:prs}).

\subsection{Energetics}

Given the spectroscopic redshift of the \FRB\ host above, we can estimate the isotropic equivalent spectral energy \citep{Petroff19}

\begin{equation}
    E_\nu=\frac{4\pi D_\text{L}^2F_\nu\Delta\nu}{(1+z)\nu},
\end{equation}

\noindent where $D_\text{L}$ is the luminosity distance to the source, $F_\nu$ is the specific fluence, and $\Delta\nu$ is the spectral extent of the burst. For the brightest (U29) and faintest (U12) burst in our sample, we obtained a spectral energy of $\sim1\times10^{30}\,\text{erg}\,\text{Hz}^{-1}$ (with $\Delta\nu\sim300$\,MHz) and $\sim6\times10^{27}\,\text{erg}\,\text{Hz}^{-1}$ (with $\Delta\nu\sim100$\,MHz), respectively. The faintest burst is only one order of magnitude more energetic than the most energetic burst from the Galactic magnetar SGR 1935+2154 \citep{CHIME20, Bochenek20}. Compared to the faintest burst from the nearby FRB 20200120E in M81, it is about four orders of magnitude more energetic \citep{Nimmo22}. Note that this energy gap could be larger if the burst has emission outside of our observing band.

\subsection{Host galaxy DM contribution}

The observed DM of \FRB\ can be separated into four components:

\begin{equation}    \text{DM}_\text{obs}=\text{DM}_\text{MW}+\text{DM}_\text{MW,halo}+\text{DM}_\text{IGM}+\text{DM}_\text{host},
\label{eq:DM}
\end{equation}

\noindent where $\text{DM}_\text{MW}$ is the contribution from the Milky Way's interstellar medium, $\text{DM}_\text{MW,halo}$ is from the Milky Way halo, $\text{DM}_\text{IGM}$ is from the intergalactic medium (IGM), and $\text{DM}_\text{host}$ is from the FRB host including its halo and any gas local to the FRB source, all in the observer's frame. We take $\text{DM}_\text{MW}=50\,\text{pc}\,\text{cm}^{-3}$ along the FRB line of sight based on the NE2001 model \citep{Cordes02} prediction (compared with $40\,\text{pc}\,\text{cm}^{-3}$ from the YMW16 model; \citealt{YMW16}), and allow for a generous uncertainty of $\pm20\,\text{pc}\,\text{cm}^{-3}$. For $\text{DM}_\text{MW,halo}$ we adopt a range between $25\,\text{pc}\,\text{cm}^{-3}$ and $80\,\text{pc}\,\text{cm}^{-3}$ \citep{Prochaska19b, Yamasaki20}. At low redshifts, we can approximate the $\text{DM}\text{--}z$ relation as a linear function \citep{Zhang18b, Pol19, Cordes22}

\begin{equation}
\begin{aligned} \langle\text{DM}_\text{IGM}\rangle\approx(855\,\text{pc}\,\text{cm}^{-3})z\left(\frac{H_0}{67.74\,\text{km}\,\text{s}^{-1}\,\text{kpc}^{-1}}\right) \\
\times\left(\frac{\Omega_\text{b}}{0.0486}\right)\left(\frac{\chi}{7/8}\right)\left(\frac{f_\text{IGM}}{0.83}\right),
\end{aligned}
\end{equation}

\noindent where $H_0$ is the Hubble constant, $\Omega_\text{b}$ is the energy density fraction
of baryons, $\chi$ is the free electron number per baryon in the Universe, and $f_\text{IGM}$ is the fraction of baryons in the IGM. The cosmological
parameters are normalised to the standard values measured by the Planck mission \citep{Ade16}, and $f_\text{IGM}$ is normalised to $\sim0.83$ \citep{Fukugita98, Li20b}. We can derive a range of values for $\text{DM}_\text{IGM}$ assuming a log-normal distribution with parameters:

\begin{equation}
    \sigma_{\ln\text{DM}_\text{IGM}}=\{\ln[1+(\sigma_{\text{DM}_\text{IGM}}/\text{DM}_\text{IGM})^2]\}^{1/2},
\end{equation}
\begin{equation}
    \mu_{\ln\text{DM}_\text{IGM}}=\ln\langle\text{DM}_\text{IGM}\rangle-\sigma_{\ln\text{DM}_\text{IGM}}^2/2,
\end{equation}

\noindent where $\sigma_{\text{DM}_\text{IGM}}=\sqrt{\langle\text{DM}_\text{IGM}\rangle\times50\,\text{pc}\,\text{cm}^{-3}}$ characterises the cosmic variance of the IGM density (e.g. \citealt{McQuinn14}). For $z=0.13$ this gives $\text{DM}_\text{IGM}=92^{+78}_{-42}\,\text{pc}\,\text{cm}^{-3}$. Altogether, we found the host-galaxy DM of \FRB\ to be $\text{DM}_\text{host}=333_{-125}^{+90}\,\text{pc}\,\text{cm}^{-3}$. Given this substantial host DM contribution and the host possibly being classified as a dwarf star-forming galaxy (see Section~\ref{sec:host}), it is worth considering the potential existence of a PRS associated with \FRB. 

The host DM could be smaller if the FRB sightline intersects foreground galaxy clusters. 
We searched for foreground galaxy clusters that could intersect the sightline of \FRB\ in the galaxy cluster catalogue from the DESI legacy imaging survey \citep{Zou21}. Within 30\,arcmin of the FRB location, we found two galaxy clusters in the foreground: one with a characteristic radius of $R_{500}=0.80$\,Mpc (defined as the radius within which the mean density is $500\times$ the critical density of the universe) and the brightest cluster galaxy (BCG) at a photometric redshift of $z=0.088$; the other with $R_{500}=0.86$\,Mpc and the BCG at a spectroscopic redshift of $z=0.091$. However, the angular separation between \FRB\ and these two clusters (25.6\,arcmin and 11.6\,arcmin, respectively) is much larger than the angular size of the clusters (7.8\,arcmin and 7.4\,arcmin, respectively). Therefore, we do not consider the foreground DM contribution to be a significant contribution in this case. Future surveys such as FLIMFLAM \citep{Lee22} may reveal more galaxies and their redshifts in the foreground of localised FRBs and allow us to better constrain the FRB host DM.

\subsection{Potential PRS association}\label{sec:prs}

\FRB\ is hyperactive and has a dwarf star-forming host galaxy and a significant host DM, similar to FRB 20121102A and FRB 20190520B. Meanwhile, it has an RM comparable to that observed for FRB 20201124A ($\sim900\,\text{rad}\,\text{m}^{-2}$), the third FRB with a PRS detection \citep{Bruni23}. All these could suggest the presence of a PRS associated with \FRB.
Assuming that the observed RM mostly originates from the persistent emission region, we can estimate the luminosity of the PRS using a simple relation \citep{Yang20, Yang22}

\begin{equation}
\begin{aligned}
L_\nu\backsimeq5.7\times10^{28}\,\text{erg}\,\text{s}^{-1}\,\text{Hz}^{-1} \\
\times\,\zeta_e\gamma_\text{th}^2\left(\frac{|\text{RM}|}{10^4\,\text{rad}\,\text{m}^{-2}}\right)\left(\frac{R}{10^{-2}\,\text{pc}}\right),
\label{eq:per}
\end{aligned}
\end{equation}

\noindent where $\zeta_e$ is the number density ratio between the relativistic (radiating synchrotron emission) and nonrelativistic (thermal and contributing to RM) electrons, $\gamma_\text{th}$ is the typical Lorentz factor of the thermal electrons, and $R$ is the size of the persistent emission region. $R$ can be constrained to $\sim c\Delta t_\text{per}\backsimeq10^{-2}$\,pc, where $t_\text{per}\sim10$\,day is the variability timescale of the PRS. $\zeta_e\gamma_\text{th}^2$ is constrained by the observed PRS luminosities of FRBs 20121102A, 20190520B and 20201124A to $\sim0.1\text{--}10$ \citep{Bruni23}.

In order to estimate the RM contribution from the FRB local environment, we can decompose the observed RM into individual components

\begin{equation}
    \text{RM}_\text{obs}=\text{RM}_\text{ion}+\text{RM}_\text{MW}+\text{RM}_\text{MW,halo}+\text{RM}_\text{IGM}+\text{RM}_\text{host},
    \label{eq:RM}
\end{equation}

\noindent where $\text{RM}_\text{ion}$ is the ionospheric contribution from the Earth’s atmosphere, and the other components are the same as defined in Eq.~\ref{eq:DM}, all in the observer's frame. Given that $\text{RM}_\text{ion}$ ($\sim\pm1\,\text{rad}\,\text{m}^{-2}$; \citealt{Sobey19}) 
is expected to be small, we ignore this term in Eq.~\ref{eq:RM}. We also ignore $\text{RM}_\text{IGM}$ as the magnetic field along the line of sight in the IGM is fairly weak ($<21$\,nG; \citealt{Ravi16}) and could undergo multiple field reversals. The Galactic RM contribution (i.e. $\text{RM}_\text{MW}+\text{RM}_\text{MW,halo}$) can be estimated using an all-sky interpolated map of the foreground MW RM contribution, which is constructed with RM measurements of radio galaxies and has a pixel scale of $\sim1.3\times10^{-2}\,\text{deg}^2$ \citep{Hutschenreuter22}. Towards \FRB, $\text{RM}_\text{MW}+\text{RM}_\text{MW,halo}=-14\pm10\,\text{rad}\,\text{m}^{-2}$. Therefore, the redshift corrected host RM is $449\pm13\,\text{rad}\,\text{m}^{-2}$. This consists of RM contributions from the FRB host galaxy and local environment. As we do not have information to distinguish between these two parts, we adopt a conservative estimate of $0\pm10\,\text{rad}\,\text{m}^{-2}$ for the FRB host based on the extragalactic RM distribution of polarised radio galaxies \citep{Vernstrom19, Carretti22}. Taking into account the above considerations, we arrived at an RM contribution from the FRB local environment of $449\pm23\,\text{rad}\,\text{m}^{-2}$. Using this value in Eq.~\ref{eq:per} we obtain a PRS luminosity of $\backsimeq[0.25\text{--}25]\times10^{27}\,\text{erg}\,\text{s}^{-1}\,\text{Hz}^{-1}$, corresponding to a flux density of $\backsimeq[0.6\text{--}60]\,\mu\text{Jy}$ at the luminosity distance of \FRB. Searching for such a faint PRS would require deep continuum observations, e.g. on an integration time of $\sim$\,hr with the MeerKAT L-band, and 
will be undertaken in a future study.

At the time of writing this manuscript, there has been a suggestion of a radio continuum counterpart associated with \FRB\ \citep{Zhang24c}. We are also conducting a study of the imaging data obtained as part of this DDT and the results will be part of a separate paper. However, given the best resolution that MeerKAT can achieve is a few arcseconds, we still need VLBI follow-up to confirm the compact nature of any potential PRS.
\section{Conclusions}

In this paper, we report the observations of \FRB\ with MeerKAT, and the detection of 62 bursts in 2\,hr of exposure, including 44 in the UHF and 18 in the L-band. This confirms the high activity of the FRB 20240114 source since its discovery by CHIME/FRB. With the voltage buffer data triggered by the brightest bursts in the L-band, we make the first arcsecond localisation of the FRB source, facilitating further follow-up observations with other telescopes. This also enables us to establish a robust host galaxy association.

We find the repeat bursts of \FRB\ are band limited with a bandwidth of $\sim210\,\text{--}\,270$\,MHz ($\sim30\%\,\text{--}\,40\%$) and show frequency downward drift with a drift rate between $\sim-0.1$ and $\sim-34\,\text{MHz}\,\text{ms}^{-1}$, similar to other repeating FRB sources. The fluences of the bursts we detected at UHF range from 0.12\,Jy\,ms to 7.02\,Jy\,ms and follow a power-law distribution with an index of $\gamma=-1.8\pm0.2$ above the 1\,Jy\,ms fluence completeness limit. The overall burst rate (including all the bursts we detected) is $44\,\text{hr}^{-1}$ and $18\,\text{hr}^{-1}$ at UHF and L-band, respectively. We find the bursts detected in the 1\,hr UHF observation approximately  follow the Poissonian repetition with a constant rate $\lambda\sim1/81\,\text{s}^{-1}$ though long-term monitoring is needed to reveal any clustering behavior.

We also investigate the polarisation properties of \FRB\ using the triggered voltage buffer data. We find most of the bursts are $\sim100\%$ linearly polarised and up to $\sim20\%$ circularly polarised. The measured PPAs show a diversity of variations across the bursts, similar to that observed in bursts detected from FRB 20180301A, suggesting a magnetospheric origin of the FRB emission \citep{Luo20}.

We identify the host galaxy of \FRB\ to be SDSS J212739.84+041945.8, which has a spectroscopic redshift of $z=0.1300\pm0.0002$. At such redshift, the spectral energy of the bursts is estimated to be in the range of $\sim6\times10^{27}\,\text{erg}\,\text{Hz}^{-1}\,\text{--}\,1\times10^{30}\,\text{erg}\,\text{Hz}^{-1}$, and the DM contributed by the host galaxy is $333_{-125}^{+90}\,\text{pc}\,\text{cm}^{-3}$. Assuming there exists a PRS associated with \FRB, we predict its flux density to be $\backsimeq[0.6\text{--}60]\,\mu\text{Jy}$ based on the simple relation between
the luminosity of the PRS and the RM \citep{Yang20, Yang22}. We encourage deep continuum observations to search for any potential PRS.

\section*{Acknowledgements}

JT would like to thank P. Weltevrede for help with the {\sc psrsalsa} software. The authors would like to thank the Director and the operators of MeerKAT and the South African Radio Astronomy Observatory (SARAO) for the prompt scheduling of the observation. 
This project has received funding from the European Research Council (ERC) under the European Union’s Horizon 2020 research and innovation programme (grant agreement No. 694745). JT and BWS acknowledge funding from an STFC Consolidated grant.
IPM acknowledges funding from an NWO Rubicon Fellowship, project number 019.221EN.019. MC acknowledges support of an Australian Research Council Discovery Early Career Research Award (project number DE220100819) funded by the Australian Government. Parts of this research were conducted by the Australian Research Council Centre of Excellence for Gravitational Wave Discovery (OzGrav), project number CE170100004. The MeerKAT telescope is operated by the South African Radio Astronomy Observatory (SARAO), which is a facility of the National Research Foundation, an agency of the Department of Science and Innovation. SARAO acknowledges the ongoing advice and calibration of GPS systems by the National Metrology Institute of South Africa (NMISA) and the time space reference systems department of the
Paris Observatory.The FBFUSE beamforming cluster was funded, installed, and operated by the Max-Planck-Institut fur Radioastronomie and the Max-Planck-Gesellschaft. 

\section*{Data availability}
The MeerKAT data underlying this paper will be shared on reasonable request to the corresponding author.



\bibliographystyle{mnras}
\bibliography{bib} 



\appendix

\section{Polarisation profiles}\label{appendix:pol}

\begin{figure*}
\centering
\includegraphics[width=.95\textwidth]{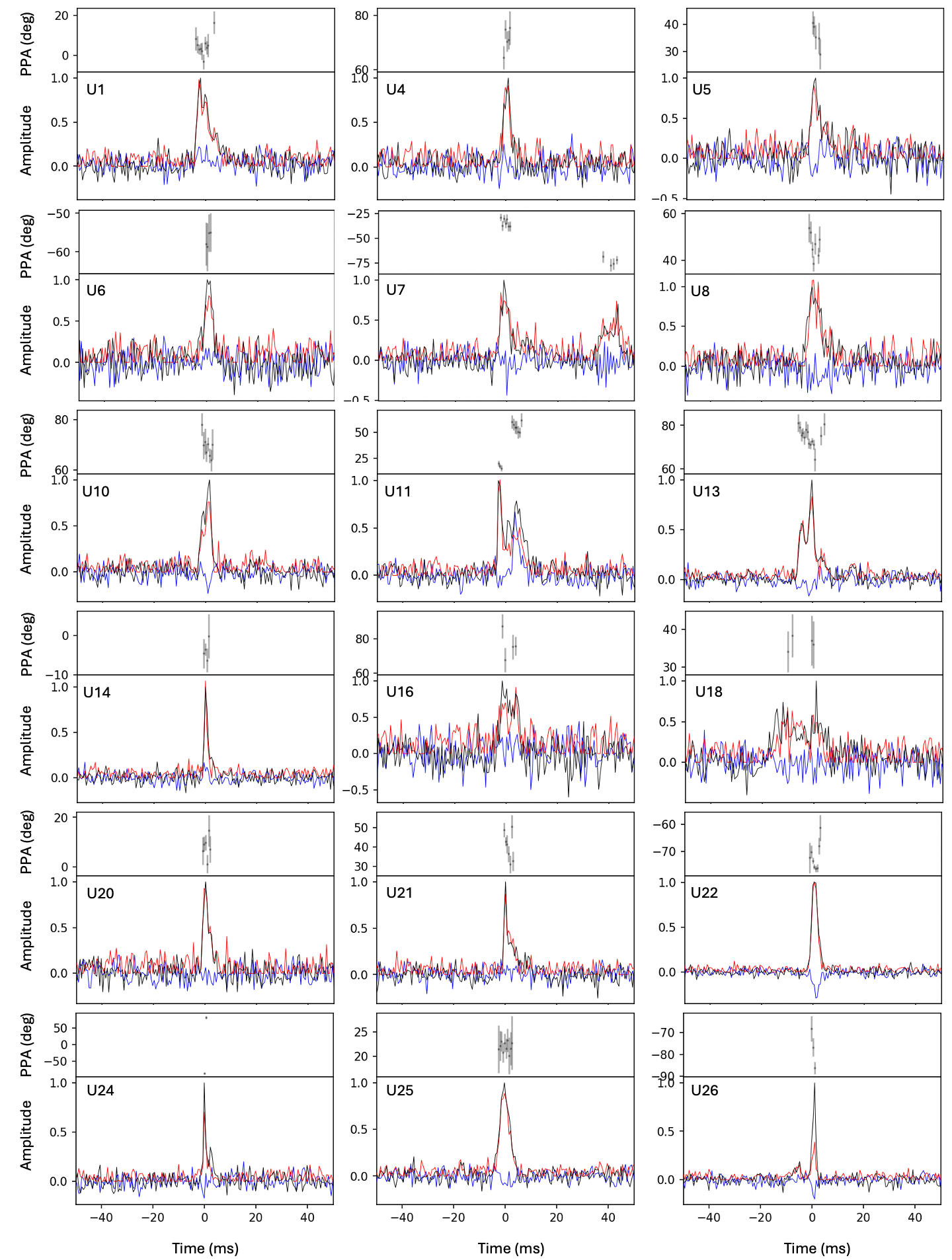}
\caption{Polarisation profiles of all the MeerKAT-detected repeat bursts from \FRB\ with polarisation data. The top of each panel shows the PPA, and the bottom shows the total intensity $I$ (black), linear polarisation $L$ (red) and circular polarisation $V$ (blue). These data are all coherently dedispersed to the DM value determined in Section~\ref{sec:DM} and derorated to the RM value in Section~\ref{sec:polarimetry}.}
\label{fig:pol_all}
\end{figure*}

In Figure~\ref{fig:pol_all} we show the polarisation profiles of all the bursts we detected in the MeerKAT observation and triggered the voltage buffer dump (see Table~\ref{tab:bursts}). The different Stokes parameters are displayed in different colors with total intensity $I$ being black, linear polarisation $L$ red and circular polarisation $V$ blue, and the PPA is displayed in the top of each panel. These polarisation data are used to calculate the linear and circular polarisation fractions for each burst, as shown in Table~\ref{tab:bursts}. Note that although bursts U44 and L9 have no triggered data, they are within the 300\,ms voltage data triggered by the next burst (see Figure~\ref{fig:bursts} for their proximity to the next burst), and thus have polarisation data. However, burst U44 is not shown here for its low S/N.

\renewcommand{\thefigure}{A\arabic{figure} (Continued.)}
\addtocounter{figure}{-1}

\begin{figure*}
\centering
\includegraphics[width=.95\textwidth]{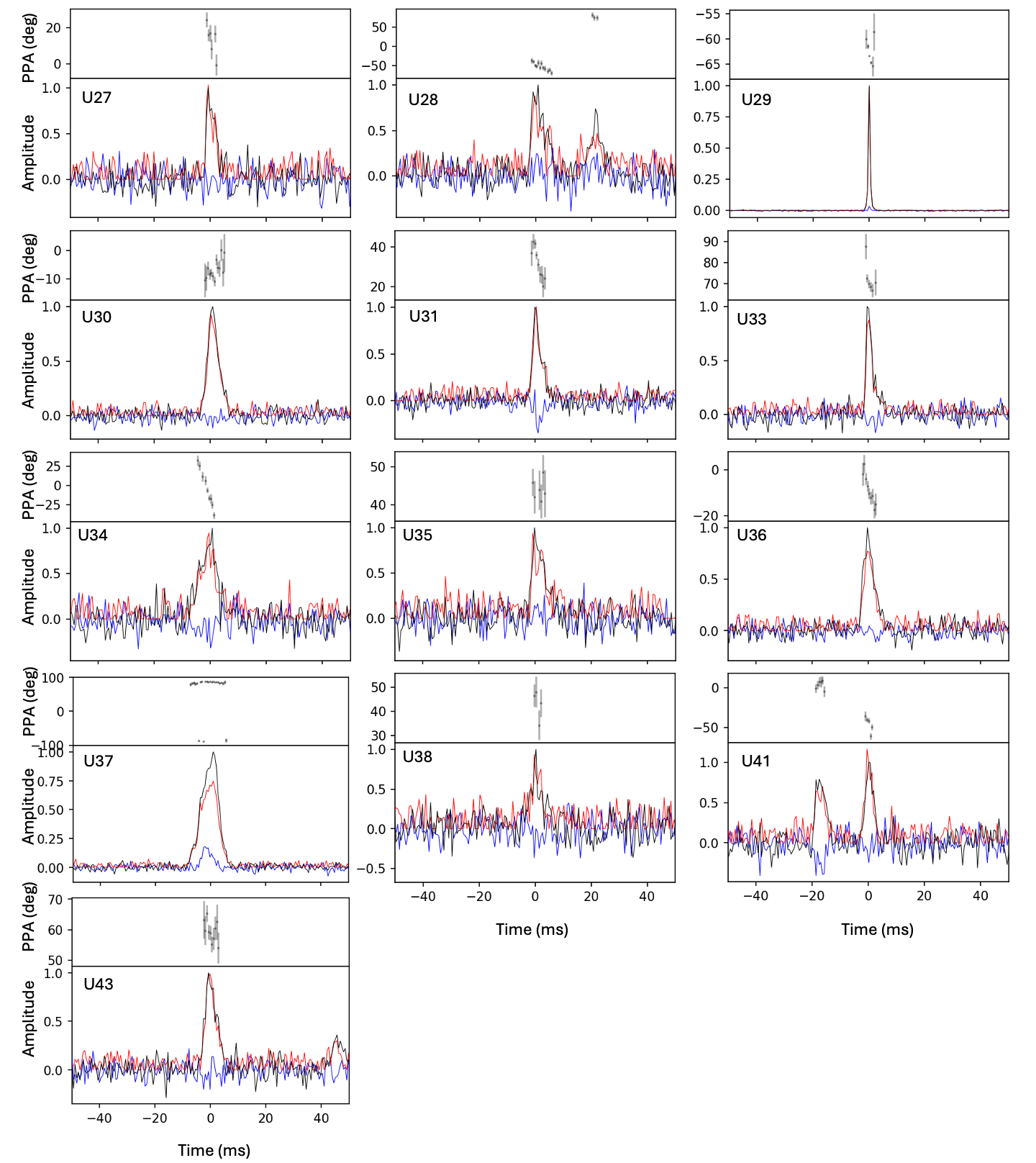}
\caption{}
\end{figure*}

\renewcommand{\thefigure}{\arabic{figure}}

\renewcommand{\thefigure}{A\arabic{figure} (Continued.)}
\addtocounter{figure}{-1}

\begin{figure*}
\centering
\includegraphics[width=.95\textwidth]{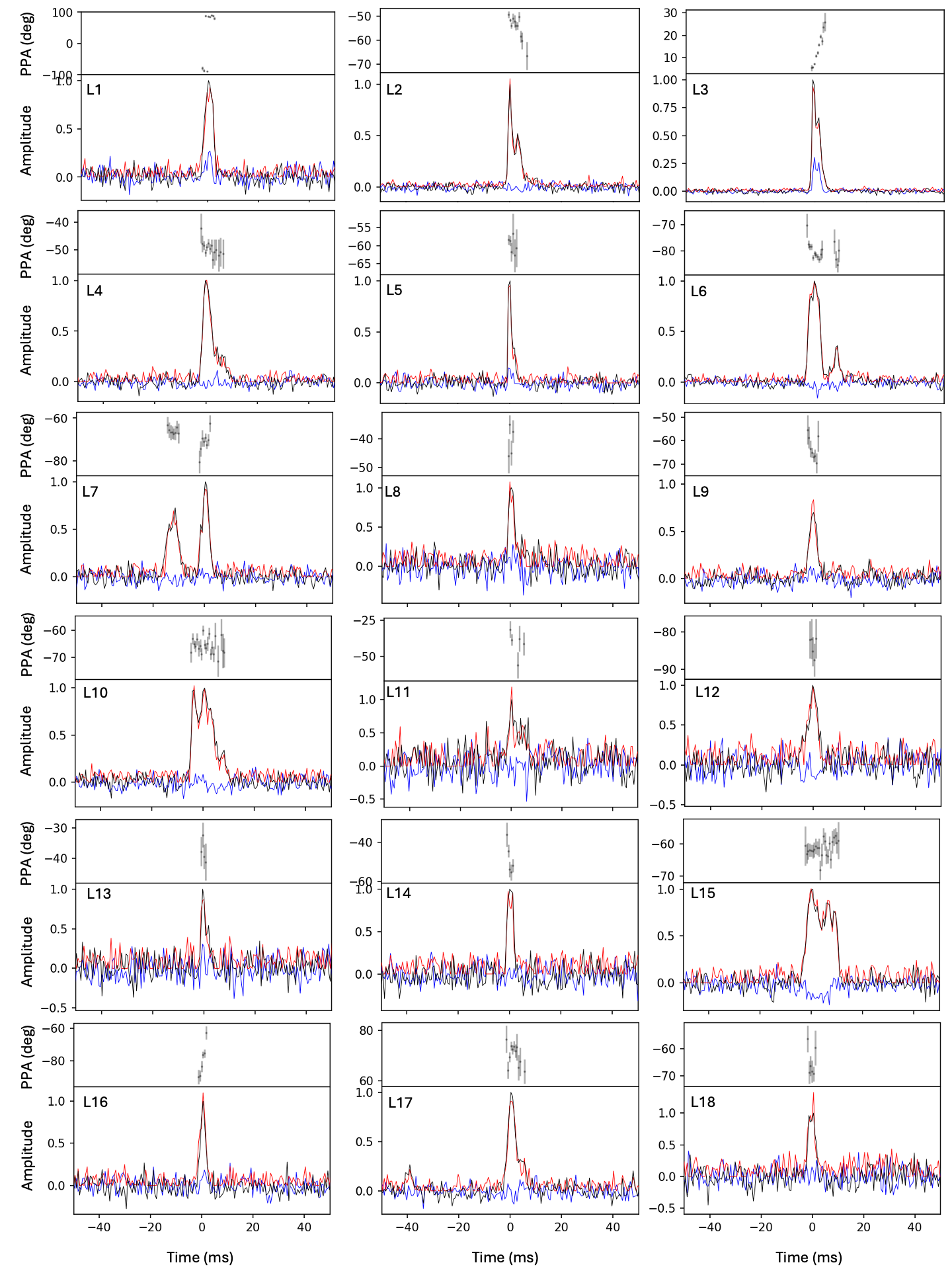}
\caption{}
\end{figure*}

\renewcommand{\thefigure}{\arabic{figure}}


\bsp	
\label{lastpage}

\end{document}